\documentclass[11pt]{article}
\usepackage{multirow}
\usepackage{geometry}
\usepackage{enumerate}

\usepackage{bbold}
\usepackage{amsmath,amssymb,amsfonts,xfrac}
\usepackage{graphicx}
\usepackage[footnotesize]{caption2}
\usepackage{mathrsfs}
\usepackage{bbm}
\usepackage{braket}
\usepackage{cancel}
\usepackage{color}
\usepackage{mathtools}
\usepackage{url}
\usepackage[colorlinks=false, citecolor=blue, urlcolor=blue]{hyperref} 
\usepackage{cleveref}

\crefformat{section}{\S#2#1#3} 
\crefformat{subsection}{\S#2#1#3}
\crefformat{subsubsection}{\S#2#1#3}

\def\beq{\begin{equation}}
\def\eeq{\end{equation}}
\def\bea{\begin{eqnarray} }
\def\eea{ \end{eqnarray} } 

\newcommand{\eq}[1]{Eq.~\eqref{#1}}
\newcommand{\Figref}[1]{Fig.~\ref{#1}}

\definecolor{orange}{rgb}{1.0, 0.3, 0.1}

\usepackage{cite}

\makeatletter
\def\bstctlcite{\@ifnextchar[{\@bstctlcite}{\@bstctlcite[@auxout]}}
\def\@bstctlcite[#1]#2{\@bsphack
 \@for\@citeb:=#2\do{%
   \edef\@citeb{\expandafter\@firstofone\@citeb}%
   \if@filesw\immediate\write\csname #1\endcsname{\string\citation{\@citeb}}\fi}%
 \@esphack}
\makeatother

\begin{document}
\bstctlcite{IEEEexample:BSTcontrol}
\vspace*{-2cm}
\begin{flushright}
KIAS-P21017
\end{flushright}
\begin{center}
{\Large\bf Swampland de Sitter Conjectures in \\ No-Scale Supergravity Models}
\end{center}

\vspace{0.5cm}
\begin{center}{
{Ida~ M.~ Rasulian}$^{1}$,
{ Mahdi~Torabian}$^1$,
and { Liliana~Velasco-Sevilla}$^{2,3}$\\
}
\end{center}

\begin{center}
{\em $^1$Department of Physics, Sharif University of Technology, Tehran 11155-9161, Iran};\\[0.2cm]
{\em $^2$ Department of Physics and Technology, University of Bergen,\\
PO Box 7803, 5020 Bergen, Norway;} \\[0.2cm]
{\em $^3$Korea Institute for Advanced Study, Seoul 02455, Korea}
\end{center}
\begin{abstract}

	It is challenging to construct explicit and controllable models that realize de Sitter solutions in string compactifications. This difficulty is the main motivation for the  \emph{Refined de Sitter Conjecture} and the \emph{Trans-Planckian Censorship Conjecture} which forbid stable de Sitter solutions but allow metastable, unstable and rolling solutions in a theory consistent with quantum gravity. 
 Inspired by this, we first study a toy de Sitter No-Scale Supergravity model and show that for particular choices of parameters it can be consistent with the Refined de Sitter Conjecture and the Trans-Planckian  Censorship Conjecture. Then we modify the model  by adding rolling dynamics and show that  the theory can become stable along the imaginary direction, where it would otherwise be unstable.  We extend the model to multi-field rolling and de Sitter fields,  finding  the parameter space where they can be compatible with the Refined de Sitter Conjecture . The modified models with rolling fields can be used to construct Quintessence models to accommodate the accelerating expansion of the Universe.

\end{abstract}
\tableofcontents
\section{Introduction and Motivation\label{eq:deSitterSec}}

In recent years, the search for de Sitter (dS)  solutions in superstring and supergravity theories has intensified {\footnote{The literature is vast, see for example  
\cite{Kachru:2003aw,Balasubramanian:2005zx,Westphal:2006tn,Rummel:2011cd,Blaback:2013fca,Cicoli:2013cha,Cicoli:2015ylx,Heckman:2018mxl,Heckman:2019dsj}.}  due to two pivotal observations. The first one is the discovery of the accelerating expansion of the Universe due to non-vanishing vacuum energy  \cite{Riess:1998cb, Riess:1998dv}. The second one is the observational support for inflationary cosmology \cite{Akrami:2018odb}.  
 According to the latter observation, it appears that the Universe underwent an early epoch of near-exponential quasi-de Sitter expansion driven by an inflaton field energy that was large in comparison with the electroweak scale of the Standard Model, but still hierarchically smaller than the Planck scale.

Nevertheless, it has been challenging to construct explicit and controllable models that realize de Sitter solutions in string compactifications \cite{Maldacena:2000mw,Hertzberg:2007wc,Covi:2008ea,deCarlos:2009fq,Wrase:2010ew,Shiu:2011zt,Green:2011cn,Dasgupta:2014pma,Kutasov:2015eba,Quigley:2015jia,Junghans:2016uvg,Andriot:2016xvq,Junghans:2016abx,Moritz:2017xto,Sethi:2017phn,Danielsson:2018ztv}. This difficulty is the main motivation for the proposal of the Refined  de Sitter Conjecture (RdSC), \cite{Ooguri:2018wrx,Obied:2018sgi} (see also \cite{Garg:2018reu}). In particular, the RdSC constrains the scalar potential of the low energy effective field theory (EFT)  and rules out the existence of dS minima, nevertheless, dS maxima are allowed. Namely, the scalar potential $V$ of a consistent EFT must satisfy either 
\bea
\label{eq:firstrdeSitter}
|\nabla V | \geq \frac{c}{M_{\rm{P}}}  \ V\, ,
\eea
or
\bea
\label{eq:hessian}
\text{min} \left(\nabla_i \nabla_j V \right) \leq  -\frac{c'}{M^2_{\rm{P}}}\ V\, ,
\eea
where the constants $c$, $c'$  are order one positive constants.   In our work we refer respectively to Eqs.~(\ref{eq:firstrdeSitter}) and ~(\ref{eq:hessian}) as the first and second de Sitter criteria. They are referred with that terminology in the literature or also with the name {\emph{conjectures}}.  Eqs.~ (\ref{eq:firstrdeSitter}) and  (\ref{eq:hessian}) imply that stable de Sitter vacua do not exist, however, unstable vacua or rolling dS solutions can exist. On the other hand,  the Trans-Planckian Censorship Conjecture (TCC) \cite{Bedroya:2019snp}, another swampland conjecture related to dS background, puts an upper bound on the lifetime, $\tau$, of a dS solution. In particular, besides unstable and rolling dS solutions, it allows for metastable dS minima with limited lifetime. If we call $H_f$,  the Hubble rate at the end of dS phase, given by $H_f \sim \sqrt{V/ 3\,  M_{\rm{P}}^2 }$, then the TCC implies
\beq 
\label{eq:TCC}
\tau < \frac{1}{H_f}\, \ln\frac{M_{\rm{P}}}{H_f}.
\eeq
Hereafter we work in Planck units, hence $ M_{\rm{P}}=1$. If the above bound, \eq{eq:TCC},  is applied to the scalar potential, $V$, then dS extrema are forbidden in asymptotic regions of moduli space; however similar to the first dS criteria, \eq{eq:firstrdeSitter}, rolling dS solutions are allowed for a fixed value of $c=\sqrt{2/3}$. On the other hand, in the bulk of moduli space, the TCC allows dS critical points (both minima and dS maxima) as long as their lifetime is bounded.

Some studies indicate that within the Swampland Programme there is a web of conjectures instead of a list of conjectures, {\it i.e.}, the conjectures are related to each other (see for instance \cite{Andriot_2020}).  As already pointed out, the TCC implies the RdSC in the asymptotic region of the moduli space. Besides, the TCC fixes the universal order one constant of the RdSC bound of \eq{eq:firstrdeSitter}. Furthermore, there is an interesting coincidence between the TCC lifetime of a dS solution and the scrambling time of  a  black hole \cite{Bedroya:2019snp}. The upper bound on the lifetime, indicates that the dS vacuum is not a thermal state. Namely, there is not enough time for the perturbations to get thermalized with the state. The TCC condition can also be seen as a bound on the growth rate of the entropy  \cite{bedroya2020sitter,Aalsma_2020,aalsma2021shocks}. The latter in particular indicates a deep connection between the TCC and a principle of quantum gravity. We recall that the dS state is a state with finite entropy and, depending on its natural scale (its Hubble rate), there is a bound on how fast it can be saturated with microstates. There are indications \cite{Palti:2019pca} that the UV/IR decoupling limit of effective field theories (without gravity) might not be working in the presence of gravity. 
 In this sense, the TCC bound offers a way to test  the IR cutoff (the Hubble scale) and the UV cutoff (the Planck scale) of a theory.   This is because the TCC is a concrete example of the UV/IR non-decoupling effects, hence offering a way to understand a theory in terms of a principle of quantum gravity. 
  All these arguments lead us to apply the TCC besides the RdSC in our present bottom-up study.

Although a priori there is no no-go theorem, it is not clear whether string theory admits dS solutions. 
The RdSC  belongs to the swampland programme that aims at determining the list of criteria an EFT must admit to be embedded in a  theory of quantum gravity. Namely, not every quantum field theory admits an ultraviolet completion when quantum gravity is considered. For bottom-up model building, the list of swampland conjectures helps to construct consistent models that can possibly descend from string theory.

In \cite{Ellis:1983ei}, and then in \cite{Roest:2015qya} for particular cases,  de Sitter no-scale models have been constructed by building de Sitter plateaus from the combination of Minkowski solutions.  These scenarios have been then subsequently and extensively applied to construct inflation models, \cite{Ellis:2018xdr,Ellis:2019hps, Ellis:2019dtx, Ellis:2020lnc}. No-scale supergravity models avoid technical problems of generic supergravity theories and thus are useful for making realistic calculable models. One may wonder if no-scale supergravity models which have dS vacua admit the swampland conjectures and thus can be derived from strings or not, in which case belonging to the swampland. Recently in  \cite{Ferrara:2019tmu}  the RdSC were re-phrased in  the context of ${\cal N}=1$ supergravity chiral multiplets. In particular for a positive potential $V>0$ the first criterion of RdSC, \eq{eq:firstrdeSitter}, implies 
\beq 
\label{eq:RdSC}
 \frac{K^{I\bar J}\, \partial_{I} V \, \partial_{\bar J}\, V} {V^2}\geq c^2/2,
\eeq
where $K^{I\bar J}$ is the inverse of the K{\"a}hler metric and $I$ runs over all scalar components of chiral superfields. In this paper we use the proposal of \cite{Ferrara:2019tmu}, which  offers an escape  to  \eq{eq:RdSC}, for  models with de Sitter vacua by modifying the theory 
with the addition of rolling dynamics. This proposal adds an exponential dependence on the rolling field to the original scalar potential, $V_0$, such that the full scalar potential becomes $\widehat V= e^{\lambda\,  \chi} \, V_0$, with  $\lambda$ constant.  Then $\widehat V$ can be consistent with \eq{eq:RdSC} (with the appropriate replacement $V \rightarrow \widehat V$) for some choice of parameters.  Using the above approach, we find that the modification of adding a rolling dynamics, opens up the region where the models survive the swamp. We also find that for some cases these theories can be used for Quintessence \cite{Agrawal:2018own}.

The paper is organized as follows: in \cref{sec:RefdeSitter} we present the statement of the Refined de Sitter 
Conjecture in the context of \cite{Ferrara:2019tmu}, then we identify the necessary conditions for the $\alpha$- supergravity/no-scale supergravity theories to satisfy \eq{eq:hessian} and the Trans-Planckian Conjecture, \eq{eq:TCC} .  
 In  \cref{sec:noscale_rollingdyn}  we study how  adding an exponential factor to the $\alpha$-supergravity/no-scale models widens the parameter space to be compatible with Eqs.~(\ref{eq:firstrdeSitter}) and ,(\ref{eq:hessian}).  In \cref{sec:nonminimal}  we study multifield models, containing more than one rolling field or more than one de Sitter field.  In  \cref{sec:Quintessence} we talk about how the models we present can be used as Quintessence models. In \cref{sec:Conclusion} we conclude.

\section{No-Scale Models with Unstable dS Vacua \label{sec:RefdeSitter} }
\subsection{No-Scale de Sitter Models}

Stable supergravity theories parameterizing a symmetric geometry  \cite{Kallosh:2013yoa,Kallosh:2014rga} have provided the basis to construct  a broad  class of superconformal inflationary models with a universal observational prediction for the spectral index of  $n_s=1-2/N $ and $r = 12/N^2$, as favored by the Planck data \cite{Aghanim:2018eyx},  where $N$ is the $N$-folds necessary for inflation.  This class of models has been generalized through the introduction of the parameter $\alpha$, inversely proportional to the curvature of the inflaton K{\"a}hler manifold in the limit of sufficiently large curvature ($\alpha$ small). As mentioned in the introduction, a construction of this kind of models as a deformation of no-scale models was constructed by building de Sitter plateaus from the combination of Minkowski solutions  \cite{Ellis:1983ei, Roest:2015qya}. These models have  been then further developed and analyzed  in \cite{Ellis:2018xdr,Ellis:2019hps, Ellis:2019dtx, Ellis:2020lnc} using the following K{\"a}hler potential and superpotential
\bea
\label{eq:noscale_gen}
K &=& -3\, \sum_{i=1}^N\, \alpha_i \, \ln(\phi_i + \bar\phi_i),\\
W &=&a\, \left(\prod_{i=1}^N\, \phi_i^{n_{i+}} -\prod_{i=1}^N\,\phi_i^{n_{i-}}\right),
\eea
where $i$ runs over $N$ no-scale chiral superfields, $\alpha_i>0$, $a$ is an arbitrary constant and 
\beq
\label{eq:ni_pm} 
n_{i\, \pm} = \frac{3}{2}\left(\alpha_i\pm\frac{r_i}{s}\right)\quad {\rm for}\quad {\textstyle\sum_{i=1}^N\, r_i^2=1},\quad  s^2={\textstyle\sum_{j=1}^N\,   \frac{r_j^2}{\alpha_j}}.
\eeq
The scalar potential
\bea
V=e^G(X-3),\quad X= G_i \, K^{i\bar \jmath }\, G_{\bar \jmath},
\eea
where  $K^{i\bar \jmath  }$ is the inverse K{\"a}hler metric and $G=K + \ln |W|^2$;  has an  extremum, $\nabla V=0$, along the real direction (i.e. $\phi_i=\bar\phi_i$) and becomes a constant in that direction \cite{Ellis:2018xdr} 
\bea
\label{eq:potVdeSconst}
V = 3\, a^2 \, 2^{2-3 \sum_{i=1}^N\, \alpha_i}.
\eea
 In \cite{Ellis:2018xdr} it was noted that in order to render the theory stable a  quartic term could be added to the K{\"a}hler potential in the following way
 \bea
 \label{eq:stImqt}
 K=-3\,  \sum_{i}^3 \, \alpha_i\,  \ln \left[\phi_i  + \overline{\phi}_i + b_i (\phi_i -\overline\phi_i)^4\right],
 \eea
 such that the imaginary direction is stabilized and hence the full potential (that is when taking into account real and imaginary terms) can remain de Sitter and hence bounded from below.

\subsection{Refined de Sitter Conjecture in No-Scale Models}
The potential of \eq{eq:potVdeSconst} is positive and hence it is  a dS solution, being constant along the real direction, cannot satisfy the first criterion of RdSC, \eq{eq:firstrdeSitter}, since the constant $c$ appearing there cannot be zero. 
In the following, we study  a simple single field no-scale model and show that the instability along the imaginary direction makes the model compatible with the second criterion of RdSC \eqref{eq:hessian}. The K{\"a}hler potential and the superpotential of the model are
\bea
\label{eq:KahlerW_dS}
K &=& -3\, \alpha\ln(\phi + \bar\phi),\nonumber\\
W &=&a\, (\phi^{n_+} - \phi^{n_-}), 
\eea
where $\alpha>0$  and $n_\pm=3/2(\alpha\pm\sqrt\alpha)$. The scalar potential is 
\bea
\label{eq:VdeSitter}
V&=& a^2 \ \left(\phi + \bar\phi \right)^{-3\, \alpha} \,
|\phi^{n_+} - \phi^{n-}|^2 \left (X -3 \right),
\eea
where
\bea
\hspace*{-1cm} && X=G_i \, K^{i\bar \jmath }\, G_{\bar \jmath } = \\
\hspace*{-1cm}  && \frac{ \left[-(\phi^{n_-}-\phi^{n_+})+ \frac{(\phi+\bar{\phi})}{3\, \alpha}\, (n_- \phi^{n_- -1}-n_+ \phi^{n_+-1}) \right] \! \left[ - (\bar{\phi}^{n_-}-\bar{\phi}^{n_+})+ \frac{(\phi+\bar{\phi})}{3\, \alpha}  (n_- \bar{\phi}^{n_- -1}-n_+ \bar{\phi}^{n_+ -1})  \right]}{3\alpha \,  \left(\phi^{{n_{-}}}- \phi^{n_{+}}\right) \left(\bar\phi^{n_{-}} - \bar\phi^{n_{+}}\right)} \, . \nonumber
\eea
We compute the eigenvalues of the squared mass terms of the scalar field $\phi$ using the following expression for the Hessian
\begin{align}
\label{eq:HessiandS}
\mathcal{H}=
\left[
\begin{array}{ll}
K^{j \bar l} \, \nabla_{i} \nabla_{ \bar l}   V \quad &   K^{l \overline m} \, g_{\overline m  i}   \, \nabla_{l} \nabla_{j}  V  \\ [10pt]
K^{i  \bar l} \, g^{j \overline m }\,  \nabla_{\bar l} \nabla_{\overline m}\,    V  \quad &  K^{i \bar l}  \, \nabla_{\bar l} \nabla_{j}   V 
\end{array}
\right],
\end{align}
where
$  \nabla_{i} \nabla_{ j}  \, V =\partial_{i}\, \partial_{j}\,   V  - \Gamma^{k} _{ij}\, \partial_{k}\,  V  $ 
and the Christoffel symbols are 
\beq 
\label{eq:Gammaor}
\Gamma^{\phi}_{\phi\phi}   = \Gamma^{\bar\phi}_{\bar\phi\bar\phi}   = -\frac{2}{\phi + \bar\phi}.
\eeq
In general $V_i$ can be simply calculated from
\bea
\label{eq:V0dergral}
\partial_{i}\,  V = V_i=e^G \left[ G_i\, (X-3) + X_i \right],
\eea
for our case it reduces to
\bea
\label{eq:derivativeOr}
\partial_{i}  V &=& -3   \left[ a^2 (-1 + 3 \alpha)  \phi^{-2 - 3\, (\sqrt{\alpha} + 
 \alpha)/2} \bar \phi^{-1 - 3\, (\sqrt{\alpha} +  \alpha)/2}  ( \phi - \bar \phi)(  \phi  +  \bar \phi )^{1 - 3  \alpha}\right],
\eea
which vanishes for 
\bea
\label{eq:conditionminor}
\phi=\bar\phi, 
\eea
making evident that the no-scale potential $V$ is flat  along the real direction. The first panel of \Figref{fig:V_comparison} shows this scalar potential for $\alpha=1$, where we compare it to the case when the imaginary direction of $\phi$ has been stabilized using \eq{eq:stImqt}, second panel of \Figref{fig:V_comparison}.
The Hessian, \eq{eq:HessiandS} for $\bar\phi=\phi$ acquires a simple form because we have just one pair of fields and hence $K^{\phi \bar \phi}\, \nabla_ {\bar \phi} \nabla_{\phi} V = K^{\bar \phi \phi}\, \nabla_ {\phi} \nabla_{\bar \phi} V $ . Also the elements above and below the diagonal are equal when $\bar\phi=\phi$,  as expected from the form of the potential  hence producing one zero eigenvalue. The non-zero eigenvalue  then is just  the trace of the Hessian {\footnote{We note that with the substitutions of  $\zeta = 2 \phi$, $\lambda_1 = 2^{-n_-} a$ and $\lambda_2 = 2^{-n_+} a$, we recover the result of  Eq.~23 of \cite{Ellis:2019dtx}.}}, corresponding to the squared mass along the imaginary direction of $\phi$
\bea
\label{eq:m2Imphi}
m^2_{\rm{Im}\ \phi}=  \frac{2^{2 - 3 \, \alpha} a^2}{ \alpha}  \, \phi^{-3 \sqrt{\alpha}} \left[\, \alpha \left(-1 + \
\, \phi^{3\,  \sqrt{\alpha}} \right)^2 - \left(1 + \, \phi^{ 3\,  \sqrt{\alpha}} \right)^2  \right].
\eea
\begin{figure}[t!]
\begin{center}
\includegraphics[width=7.8cm]{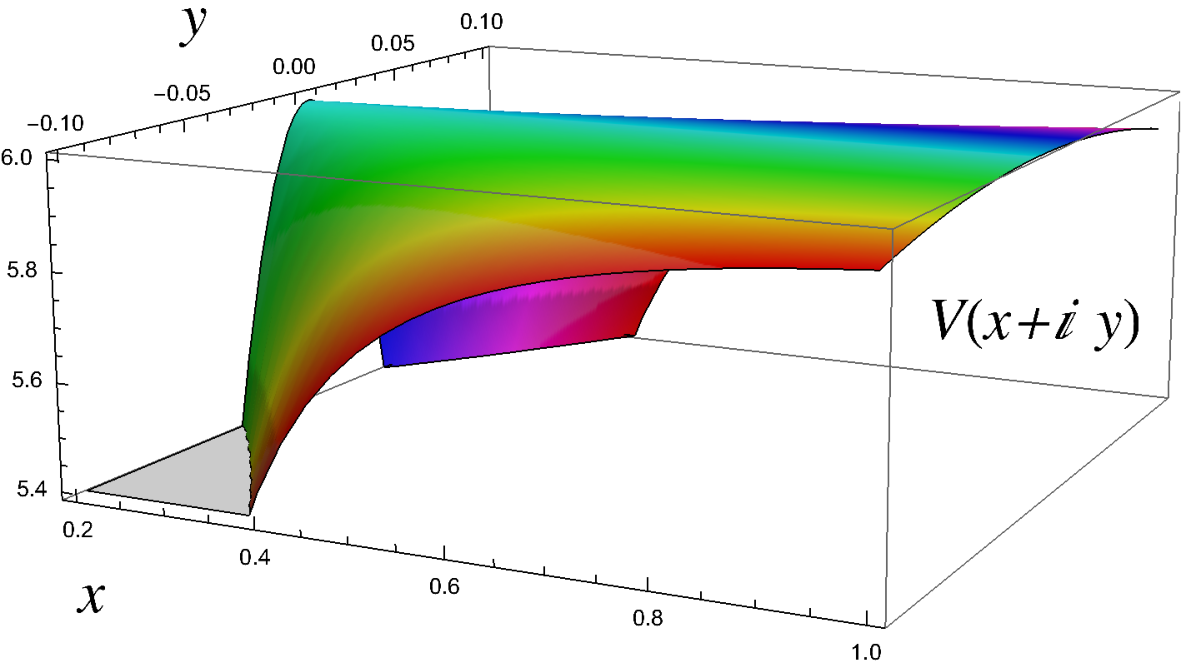}
\hspace*{6mm}
\vspace*{1.2mm}
\includegraphics[width=6.2cm]{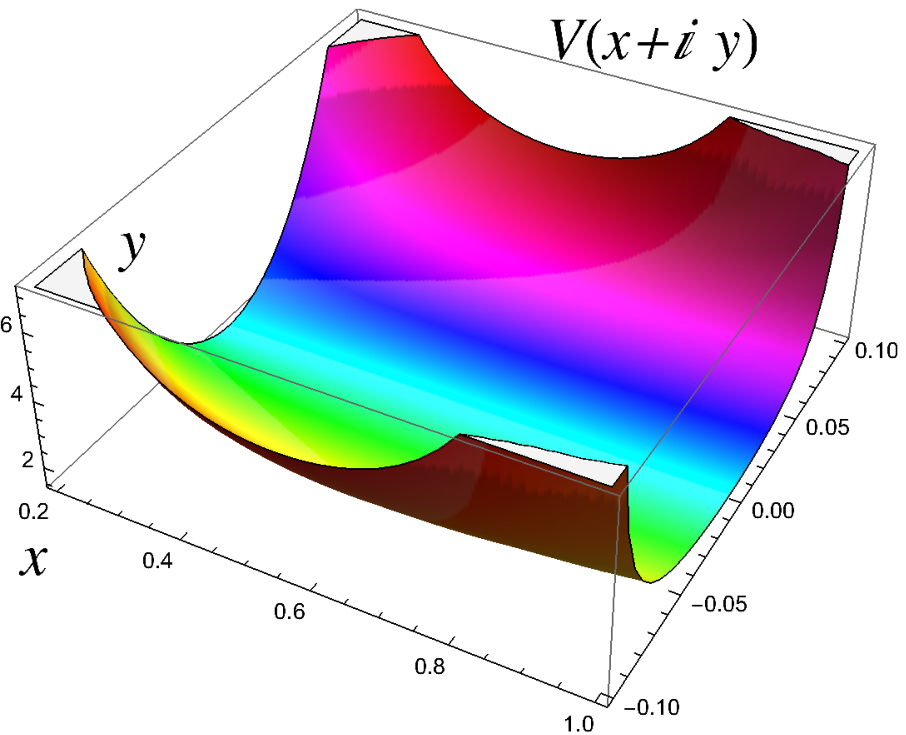}
\includegraphics[width=7.8cm]{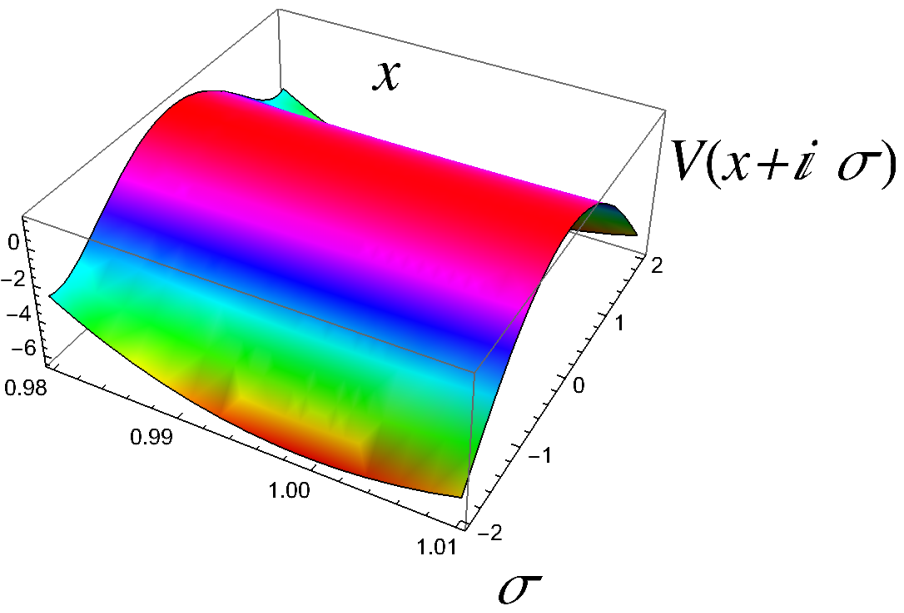}
\end{center}
\vspace*{-8mm}
\caption{First row: comparison of the scalar potential, $V(\phi)$  (left) of \eq{eq:VdeSitter}, to the potential  $V(\phi)$  but with  $K=-3 \sum_{i}^3 \, \alpha_i \ln \left(\phi_i  + \overline{\phi}_i + b_i (\phi_i -\overline\phi_i)^4\right)$, \eq{eq:stImqt}, which stabilizes the imaginary direction. Second row:  the potential $V(x + i\, \sigma)$ for $x$ and the canonical field $\sigma$. The direction $x$ gets fixed to $b$ in order to work with a canonical field, $\sigma$. For all cases we used  $\alpha=1$.
 \label{fig:V_comparison}}
\end{figure}
It is well known that for $\alpha\leq1$ there is an instability for any 
field value (e.g. \cite{Ellis:2019dtx}), while for $\alpha>1$ the non-zero eigenvalue is negative for
\bea
\label{bound1}
\left[\frac{\sqrt{\alpha}-1}{1+\sqrt{\alpha}}\right]^{\frac{1}{3\sqrt{\alpha}}}<\phi<\left[\frac{1+\sqrt{\alpha}}{\sqrt{\alpha}-1}\right]^{\frac{1}{3\sqrt{\alpha}}} .
\eea
In order to turn the imaginary component into canonical form, we need to stabilize the flat real direction.
It can be done, for instance, through modifying the K{\"a}hler potential (see \cite{Ellis:2020lnc} for a review) of the model \eq{eq:KahlerW_dS} by
\bea
\label{eq:Kahler_quartic_realterm}
K &=& -3\, \alpha\, \ln\left[\phi + \bar\phi + \, \frac{\left(\phi + \bar\phi- 2 \, b\right)^4}{L} \right],
\eea
which stabilizes the real component to $\bar\phi=\phi=b$,  and provides a way to work only with one canonical field, $\sigma$, proportional to the imaginary direction (see Appendix \ref{app:cannorm}): 
\bea
\sigma=\sqrt{\frac{3\, \alpha}{ 2\, b^2}}\, y.
\eea
The theory of \eq{eq:KahlerW_dS} can be extended to include more fields and stabilize the real directions with the same procedure.  With the addition of \eq{eq:Kahler_quartic_realterm} the scalar potential  has the same form of  \eq{eq:VdeSitter}. Note that when fixing the real part to $b$ all the dependence inversely proportional to $L$ in \eq{eq:Kahler_quartic_realterm} vanishes from the potential and their subsequent derivatives, as can be seen from  \eq{eq:VdeSitter}, the form of the first derivative, \eq{eq:derivativeOr}, and of the second derivatives  of $V$, 
\bea
\partial_i \, \partial_{\bar\jmath} \, V = V_{i\, \bar\jmath} &=& e^G\,  \left[ G_{\bar\jmath} \left(G_i \, (X-3) + X_i \right)  +   G_{i\, \bar\jmath } (X-3) +  X_{i\, \bar\jmath} + G_i \, X_{\bar\jmath}          \right]\,,\nonumber\\
\partial_i \, \partial_{j} \, V = V_{i\, j } &=& e^G\, \left[ G_{j} \left(G_i \, (X-3) + X_i \right) + G_{i\, j } (X-3) + X_{ij} + G_i \, X_j \right].
\eea 
It is clear from these expressions that up to the second derivatives in %
 $\phi$ or $\bar\phi$,  $G$ (e.g. $G_{ij}$) and X preserve the property of the K{\"a}hler potential in  \eq{eq:Kahler_quartic_realterm} to reduce to \eq{eq:KahlerW_dS} when $\bar\phi=\phi=b$. In the third panel of \Figref{fig:V_comparison} we compare $V$ as a function of $V(x+ i \sigma)$ to the model described by \eq{W-case3}
where the real direction has not been fixed (first panel) and to the case where the imaginary direction has been stabilized (left panel). We have chosen to plot $V(x+ i \sigma)$ as we can clearly see the behaviour when $\phi=\bar\phi=b$. In \Figref{fig:V-single} we plot the potential $V$ as a function of $\sigma$ for values of $\alpha$ which are perfect squares modulo 9, where we can see the transitions from maxima to minima: values for   $\alpha \leq 1 $ give de Sitter maxima, while values for $\alpha > 1 $ de Sitter minima.

For the case of having $\alpha=\frac{n^2}{9}$, $n$ an integer,  using \eq{eq:m2Imphi}, then the second criterion of the RdSC,  \eq{eq:hessian}, can be written as $-m^2_{\rm Im[ \phi]}/V -c' \geq 0$ or equivalently 
$\left[\frac{9}{n^2}-1\right] b^{2 n}+\left[\frac{18}{n^2}+2-3c'\right]\, b^n +\frac{9}{n^2}-1\geq 0$.  For $n>3$ we find that this bound can be satisfied for $b$ such that  $a_-<b^n<a_+$ with $a_{\pm}=\frac{9+n^2-3c' n^2/2\, \pm \, n^2\, \sqrt{\left(2-3c'/2\right)\left(18/n^2-3c'/2\right)}}{n^2-9}$ when $c'\leq \frac{12}{n^2} < \frac{4}{3}$. For $n<3$ instead  it can be satisfied for all values of $b$ if $\frac{4}{3}\leq c' \leq \frac{12}{n^2}$.  Finally for the special case $n=3$, the bound is satisfied for all values of $b$ if  $c'\leq \frac{4}{3}$.

\begin{figure}[t!]
\begin{center}
\includegraphics[width=7.6cm]{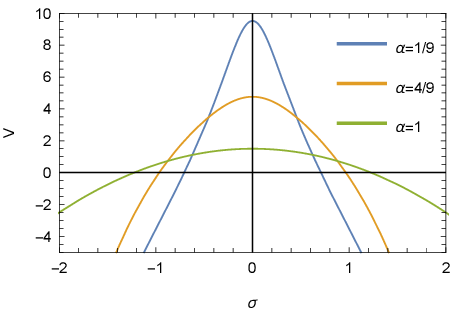}
\hspace*{2mm}
\includegraphics[width=7.6cm]{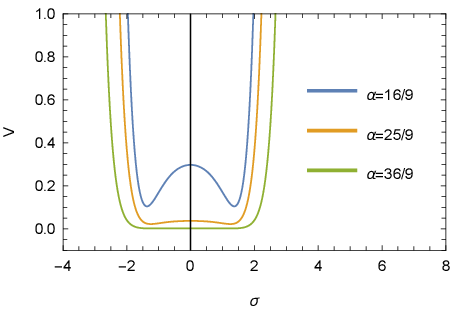}
\end{center}
\vspace*{-8mm}
\caption{The scalar potential, $V(\sigma) $, of \eq{eq:VdeSitter} with the K{\"a}hler potential of \eq{eq:Kahler_quartic_realterm} with the stabilization for $\langle {\rm{Re}} \, [\phi] \rangle = b=1$, as a function of the canonical field $\sigma=\sqrt{3\, \alpha/ 2}\  y/b$. We have chosen values of $\alpha$ which are perfect squares modulo 9. Values for   $\alpha \leq 1 $, left,  give de Sitter maxima, while values for $\alpha >1$, right, can admit de Sitter minima. \label{fig:V-single}}
\end{figure}

\subsection{Trans-Planckian Conjecture in No-Scale de Sitter Models}
The TCC implies that a de Sitter phase has a limited lifetime. In the case of a potential with de Sitter maximum, the field can sit on that critical point if the 
negative curvature, second derivative of the scalar potential, is bounded from below. Specifically, close to a local maximum if there is field range 
$0<\sigma<\Delta\sigma$ where  $|V''|\leq |V''|_{\rm max}$,    defining $V'' \equiv \partial^2_\sigma V$,  the TCC is satisfied if either of the following conditions are satisfied  \cite{Bedroya:2019snp}
\bea
\Delta\sigma &<& \frac{\frac{2}{3\sqrt{\pi}}\left(V_0 \, V_{\rm{min}}\right)^{3/4}\, \ln^{1/2}\sqrt{\frac{3}{V_{\rm{min}}}}}{\frac{2}{3}V(\sigma)- |V''|_{\rm{max}}\, \ln^2\sqrt{\frac{3}{V_{\rm{min}}}}},\label{eq:Deltasigma}\\
\frac{|V''|_{\rm{max}}}{V_{\rm{min}}} &\geq& \frac{2}{3}\frac{1}{\ln^2\sqrt{\frac{3}{V_{\rm{min}}}}}.
\eea

In the above equations  $|V''|_{\max}$, is the maximum curvature and $V_{\rm{min}}$ is the minimum value for the potential in that field range. The second inequality is similar to the second criterion, RdSC \eqref{eq:hessian}, with a prediction for $c'$ up to a logarithmic correction and thus it is a milder constraint. For the dS no-scale model we are considering and for the single field range, the  inequality \eq{eq:Deltasigma} implies that 
\bea
\label{TCC-single}
\frac{2}{3\sqrt{\pi}}\left(V_0 V_{\rm min}\right)^{3/4}\ln^{1/2}\sqrt{\frac{3}{V_{\rm{min}}}}-\left[\frac{2}{3}V_{\rm min}-|V''|_{\rm{max}}\ln^2\sqrt{\frac{3}{V_{\rm{min}}}}\right]\sigma >0.
\eea
For $\alpha\leq 1$ the expression in the square bracket is negative and  the bound is satisfied  for any value of $\sigma$. In order to see that, we note
$V_{\rm min}\leq V_0$ and $|V''|_{\rm max}\geq |V''_0|$, therefore
\bea 
\frac{2}{3}V_{\rm{min}}-|V''|_{\rm max}\ln^2\sqrt{\frac{3}{V_{\rm{min}}}} \!\!&<&\!\! \frac{2}{3}V_{0}-|V''_{0}|\,\ln^2\sqrt{\frac{3}{V_{0}}} \\ &= &2^{3-3\alpha}a^2 \left[1-\frac{b^{-3\sqrt{\alpha}}}{2\alpha}\left((1+b^{3\sqrt{\alpha}})^2-\alpha(1-b^{3\sqrt{\alpha}})^2\right)\ln^2(2^{1-\frac{3\alpha}{2}}a)\right]\,
,\nonumber
\eea
where in the second line we have used \eq{eq:m2Imphi} which can be written in the form
\bea
V''|_{\sigma=0}= -2^{2-3\alpha}a^2\frac{b^{-3\sqrt{\alpha}}}{\alpha}\left[(1+b^{3\sqrt{\alpha}})^2-\alpha(1-b^{3\sqrt{\alpha}})^2\right].
\eea
For $\alpha<1$ the above expression is negative for values of $b$ that make the potential unstable. For $\alpha=1$, the above expression is negative for $a<\sqrt 2e^{\frac{-1}{\sqrt2}}$ or  $a>\sqrt 2e^{\frac{1}{\sqrt2}}$  
 independently of $b$, as can be seen from the left panel of ~\Figref{fig:tccplot}.
We recall that for all phenomenological purposes $a\ll1$.  For $\alpha>1$, the analytical computation is involved as can be seen from the right panel of ~\Figref{fig:tccplot}. Numerical  inspection then give us for this case the values of $\sigma$ and $b$ that are compatible with the TCC bound, \eq{TCC-single}.
\begin{figure}[h]\begin{center}
\includegraphics[width=7cm]{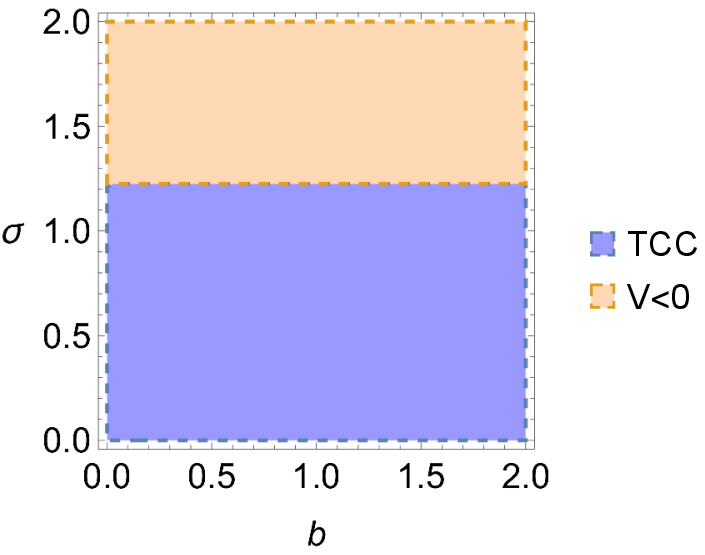}
\hspace*{5mm}
\includegraphics[width=7cm]{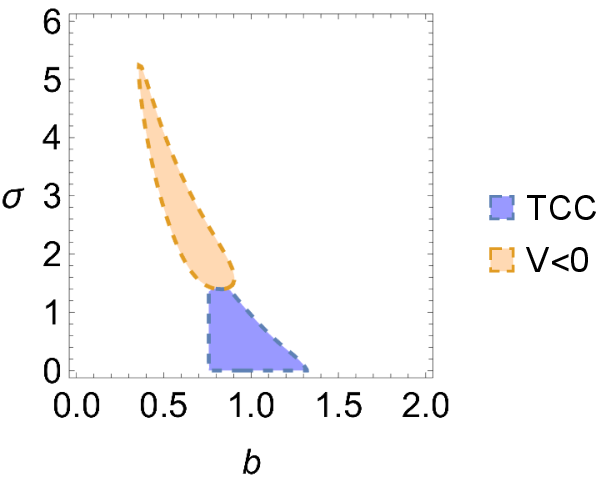}
\caption{ 
Evaluation of the TCC for $\alpha=1$, left plot, and $\alpha=25/9$, right plot. The blue regions indicate values of  $b$ and $\sigma$ which are compatible with the TCC bound when $V>0$. The orange region indicates values for which $V<0$. For both plots we have used $a=0.1$. Note that for the case of  $\alpha=1$ the value of $b$ can extend for all values along the real line.
\label{fig:tccplot}}
\end{center}\end{figure}
The TCC conjecture also allows the existence of de Sitter potentials with metastable minima as long as  these minima  are not positive or if positive by allowing quantum tunnelling.  This allows sufficiently short-lived transient quasi-dS like phases. We note that potentials with negative minima could be constructed from the kind of models of the right-side panel of \Figref{fig:tccplot}, with addition of other fields that render the minima negative.

\section{No-Scale Models with Rolling Dynamics \label{sec:noscale_rollingdyn}} 
 In this section, we  first add a rolling field to our minimal model,  then  we obtain the conditions for finding extrema, point out in which cases there can be a stable minimum, without the addition of a quartic term in the K{\"a}hler potential. Then we point out in which cases instead we can be in agreement with the first criterion of the RdSC, \eq{eq:firstrdeSitter}.
 
Following the approach proposed in \cite{Ferrara:2019tmu}, we modify the K{\"a}hler potential and the superpotential of our basic dS no-scale model as follows
\bea 
\widehat K &=& K- \sum_{m=1}^N\, q_m \ln(\chi_m + \bar\chi_m),\\
\widehat W &=& W\, \prod_{m=1}^M\, \left(c_{0,m}+c_{1,m}\, \chi_m^{-{\sfrac{p_{m,1}}{2}}}+c^2_m\, \chi_m^{-{\sfrac{p_{m,2}}{2}}} + \cdots\right).
\eea
Here $m$ runs over the number of rolling chiral superfields, $q_m>0$, $c_{k,m}$ are arbitrary constants and  the powers $p_{m,k}$ satisfy $p_{m,k}<0$. The function invariant under K{\"a}hler transformations $\widehat G= \widehat K + \ln\ \left|\widehat W\right|^2$ of the modified model is given by
\beq 
\label{eq:Gtotal}
\widehat G=G + \sum_{m=1}^M\, \left(\ln\frac{|c^0_m+c^1_m\,  \chi_m^{-{\sfrac{p_{m}^1}{2}}}+c^2_m\, \chi_m^{-{\sfrac{p_{m}^2}{2}}} + \cdots|^2}{(\chi_m + \bar\chi_m)^{q_m}}\right) \equiv G+ \widetilde G,
\eeq 
where additional chiral superfields $\chi_m$ are included to provide an escape to the RdSC for positive scalar potentials, \eq{eq:RdSC}. The addition of these chiral superfields solves at the same time the instability of $K$ at Im $\phi$=0, for which V has a maximum, \eq{eq:noscale_gen} without additional quartic terms.  The scalar potential for this model can be written then  as follows
\bea
\label{modified-pot}  
\widehat V &=& e^{\widehat G}
\left(\sum_{I=1,\bar J=1}^{N+M,N+M} \widehat G_I\, \widehat K^{I \bar J}\, \widehat G_{\bar J}-3\right)\nonumber\\
& = & e^{\widehat  G}\left(\sum_{i=1,\bar\jmath=1}^{N,N} \widehat G_{ i}\, \widehat K^{i \bar\jmath}\, \widehat G_{\bar \jmath}-3\right)+e^{\widehat G}\left(\sum_{m=1,\bar n=1}^{M,M} \widehat G_{m}\, \widehat K^{m\bar n} \, \widehat G_{\bar n}  \right) \nonumber\\
&=& e^{\widetilde G}\, \left(V+ \widehat X\, {\rm{e}}^{G} \right),
\eea
where $I=i,m$ runs over all chiral superfields, indices $i,j=1,\hdots, N$ are for the de Sitter fields, indices $m,n=1,\hdots, M$ are devoted to the rolling fields, $\widehat K^{I\bar J}$ is the inverse of the K{\"a}hler metric (which is block diagonal) 
$\widehat K_{I\bar J} = \partial_I\partial_{\bar J}\, \widehat K$ and $\widehat G_I = \partial_I \, \widehat G$, analogously for its Hermitian conjugate. 
The parameter $\widehat{X}$ is defined as
\beq 
\label{eq:definitionX}
\widehat{X} \equiv \widehat G_m\, \widehat  K^{m\bar n}\, \widehat  G_{\bar n} = \widetilde G_m\, \widehat K^{m\bar n}\, \widetilde G_{\bar n},
\eeq
where the expression after the equal sign follows from the decomposition in \eq{eq:Gtotal} and $\widehat G_m= \widetilde G_m$ since only $\widetilde G$ depends on the ``rolling" fields $\chi$. 
For the rest of this section we assume just one rolling field $\chi$. The potential $\widehat V$ is minimal in the $\chi$ plane for $\rm{Im}\, [\chi]=0$, therefore we consider only the real part of the scalar field
\bea
\gamma=\left. \widehat{X} \right|_{\widehat{V}_{\rm{Re}\, [\chi]}}.
\eea
The extremization condition of the new scalar  potential,  $\widehat{V}$, is
\bea
\label{eq:stabilityreq} 
\partial_{i}\, \widehat V= e^{\widetilde G}\, \left( \partial_{i}\, V  + \gamma \ \partial_{i}\, e^{G}\right)=
e^{\widetilde G}\, \left( \partial_{i}\, V  + \gamma \ \partial_{i}\, (e^{K} |W|^2)\right)=0.
\eea 
This equation can be satisfied in two ways.  One way is requiring both terms to cancel independently,  since $ \partial_{\bar i} V =0$  and $\partial_{\bar\imath}\, V=0$ are the conditions for the minimum of the original theory this can be easily satisfied and hence the term  $ \partial_{i}\, (e^{K} |W|^2)$ should be satisfied independently. The other way is to achieve a cancellation with both terms.

In the case of finding solutions where $W=0$, the solutions for $\partial_{i}\, V $,  \eq{eq:V0dergral}, can be written as follows
\bea
\label{eq:V0min} 
\partial_{i}\,  V=e^K\, \overline W_{\bar\jmath} \left[K^{\bar\jmath l} \left(W_l\, K_i +W_i\, K_l+W_{li} \right) + K^{\bar\jmath l}\, _i W_l\right]=0.
\eea
When the addition of the rolling field does not change the minima of the no-scale field in the supersymmetry breaking vacuum,  then the no-scale field  is stabilized and the only dynamics comes from the rolling field, as it has been emphasized in \cite{Ferrara:2019tmu}. 
For the second term in \eq{eq:stabilityreq}  we have
\bea
\label{eq:derivativeAp}
\partial_i \left(e^{K} |W|^2\right)=e^K\ \overline{W }\left(W K_i + W_i \right),
\eea
for  which we have two solutions for a vanishing term
\bea
\label{eq:twosolsvan}
W=0\quad \text{or} \quad
D_i\, W=K_i+ \frac{W_i}{W}=0,
\eea
analogously for their hermitian counterparts. The first solution in \eq{eq:twosolsvan} can be satisfied in general, while the second solution just holds for certain values of $\alpha$. 
We choose the first one, so we have  
 at the extrema
\bea
\label{eq:min_Adterm}
\prod_{i=1}^N \, \phi_i^{n_{i+}} &=& \prod_{i=1}^N\, \phi_i^{n_{i-}},\nonumber\\
\prod_{i=1}^N \, \overline\phi_i^{n_{i+}} &=& \prod_{i=1}^N\,, \overline\phi_i^{n_{i-}},
 \eea
where non trivial solutions (for which $n_{i+} \neq n_{i-}$) exist, as we will see in the next sections, hence
\bea
\label{eq:derVhat1}
&&\left. \partial_{\bar\imath} \left(e^{K} |W|^2\right) \right|_{\prod_{i=1}^N\, \overline\phi_i^{n_i+} = \prod_{i=1}^N\, \overline\phi_i^{n_i-}}=0,\nonumber\\
&&\left. \partial_i \left(e^{K} |W|^2\right) \right|_{\prod_{i=1}^N\, \phi_i^{n_i+} = \prod_{i=1}^N\, \phi_i^{n_i-}}=0.
\eea
Finally, we note that the evaluation of $\widehat V$ at the extrema for the no-scale fields implies
\beq 
\label{eq:conditiongamma}
\left. \ \frac{K^{I\bar J}\, \partial_{I} \widehat V \, \partial_{\bar J}\, \widehat V}{\widehat V^2}=  
\ \frac{\widehat K^{m\bar n}\partial_{m}\, \widehat  V\partial_{\bar n}\, \widehat V } {\widehat V^2} \right|_{\rm{\widehat{V}_{\rm{ext.}}}}=  
 \gamma\,  ,
\eeq
where we have used $\partial_{i}\, \widehat V=0$. We observe that the rolling behavior in this no-scale supergravity puts the model out of the swamp given that $2\gamma \geq c^2$. It is interesting to note, as the superpotential is vanishing in the vacuum, that we get a supersymmetry breaking vacuum with positive vacuum energy for any value of $\gamma$. This can be seen by expanding the scalar potential $\widehat V$ into pieces of the original model and the pieces coming from the rolling field
 \bea
\widehat V={e}^{\widetilde G}\, {e}^K |W|^2 \,  \left( X-3 +\gamma  \right) =  {e}^{\widetilde G} \, e^K \, \left(\sum_{i=1}^N\, \left| D_i\, W \right|^2  + (\gamma-3) |W|^2  \right),
\eea
from which it follows that at the extrema of the complete model the dependence on $\gamma$ disappears when $W=0$
\bea
\widehat V= \left. e^{\widetilde G}\, \left(\, e^K\, \sum_{i=1}^N\, |D_i\, W|^2\, \right)\right|_{\rm ext.} .
\eea
\subsection{The Minimal Model}
In this section we study the minimal model including a no-scale chiral superfield plus a rolling superfield. 
\subsubsection{Existence of Extrema}
We start with the simplest example including one no-scale chiral superfield  and one rolling chiral superfield. The K{\"a}hler and the superpotential are as follows
\bea
\label{W-case1}
\widehat K &=& -3 \, \alpha\, \ln (\phi + \bar{\phi})- q\, \ln (\chi + \bar{\chi}),\nonumber \\
\widehat W&=& a \left(\phi^{n_+}- \phi^{n_{-}}  \right)\, \chi^{-p/2}, 
\eea
where $\alpha,q>0, p<0$, $a$ is an arbitrary constant and $n_{\pm}= 3/2(\alpha\pm \sqrt\alpha)$. The  full scalar potential, $\widehat{V}$, is
as in \eq{modified-pot}, $V$, the no-scale potential is given in \eq{eq:VdeSitter} and $\widetilde G$ is 
\bea
\label{eq:G1m}
{ \widetilde G} &=& \ln\left[({2\rm Re}\, [\chi])^{-q}\, |\chi|^{-p}\right],
\eea
For the second term \eq{eq:stabilityreq} the extremization conditions of \eq{eq:derVhat1}, for the non-trivial solution (that is $n_{i-} \neq n_{i+}$), and for an arbitrary value of $\alpha$ reduce to 
\beq
\label{sol-case1}
\left.\phi\right|_{\rm ext.}=\left.\bar\phi\right|_{\rm ext.}= 1, 
\eeq
which is in agreement with the condition for finding the extrema of the original theory, \eq{eq:conditionminor}, but restricts the set of possible values for the extrema. If we instead consider the K{\"a}hler potential of \eq{eq:Kahler_quartic_realterm}, this can automatically fix $\phi=\bar\phi=1$ by choosing $b=1$ as explained in \cref{sec:RefdeSitter}.

\subsubsection{Stability Condition}
We study the stability of the vacuum by considering the Hessian matrix 
\bea
\label{eq:HessiandeSR}
\mathcal{H}=
\left[
\begin{array}{ll}
K^{J \bar L} \, \nabla_{I} \nabla_{ \bar L}   \widehat V \quad &   K^{L \overline M} \, g_{\overline M  I}   \, \nabla_{L} \nabla_{J}  \widehat  V  \\ [10pt]
K^{I  \bar L} \, g^{J \overline M }\,  \nabla_{\bar L} \nabla_{\overline M}\,  \widehat  V  \quad &  K^{I \bar L}  \, \nabla_{\bar L} \nabla_{J}  \widehat  V \, ,
\end{array}
\right]\, ,
\eea
where $I=\phi,\chi$ and $ \widehat  V_{IJ}=\nabla_{I}\, \nabla_{J}\, \widehat  V = \partial_{I}\, \partial_{J}  \widehat V  - \Gamma^{K} _{IJ}\, \partial_{K}\, \widehat   V  $. 
For the no-scale extremization where $\partial_\phi V=0$ and $e^G\sim |W|^2=0$ we find that
\bea
\left.\widehat V_{\phi\chi}\,\right|_{\rm min}=\left(\partial_\chi \, e^{\widetilde G}\right)\, \left(  \partial_\phi V \right)+ \gamma \, e^{\widetilde G}\, e^G (G_\phi\, G_\chi )=0,
\eea
similarly for any other mixed component and its Hermitian conjugate. The no-scale components do not mix with the rolling field, thus the mixed Christoffel symbols are vanishing. Therefore, we can write the Hessian matrix as 
\bea
\left.{\mathcal{H}} \right|_{\phi=\bar\phi=1}=
\left.
\begin{bmatrix}
 \widehat K^{\phi \bar{ \phi}} \, \widehat V_{ \phi \bar{ \phi}} & 0 &  \widehat K^{\phi \bar{ \phi}}  \, g_{ \bar{ \phi} \phi}\, \widehat V_{ \phi  \phi} & 0 \\
0&  \widehat K^{\chi \bar{ \chi}}\,  \widehat V_{ \chi \bar{ \chi}} & 0 &  \widehat K^{\chi \bar{ \chi}}\, g_{ \bar{ \chi} \chi}\,  \widehat  V_{ \chi  \chi} \\
 \widehat K^{\phi \bar{ \phi}}\, g^{ \bar{ \phi} \phi}\,  \widehat V_{\bar{ \phi} \bar{ \phi}} & 0 &  \widehat K^{\phi \bar{ \phi}}  \widehat V_{\bar{ \phi} \phi} & 0 \\
0 &  \widehat K^{\chi \bar{ \chi}}\, g^{ \bar{ \chi} \chi}\, \widehat V_{\bar{ \chi}\bar{ \chi}} & 0 &  \widehat K^{\chi \bar{ \chi}}\, \widehat  V_{\bar{ \chi} \chi}
\end{bmatrix}
\right|_{\phi=\bar\phi=1},
\eea
from which we can see that it is equivalent to look for the eigenvalues, which need to be positive for the stability of the scalar potential, of the following block-diagonal matrix
\bea
\label{eq:blockdiagonalH}
{\mathcal{H}}= \left[
\begin{array}{cc}
\mathcal{H}_\phi  & 0 \\
 0  & \mathcal{H}_\chi
\end{array}
\right].
\eea
Each of the blocks in \eq{eq:blockdiagonalH} contain only information from one field:
\bea
\label{eq:Hessians}
\rm{det}\left[ \mathcal{H}_\phi \right]= 
\rm{det}\left[ K^{\phi \bar{ \phi}}  \, \begin{bmatrix}
\widehat  V_{ \phi \bar{ \phi}} & \widehat   V_{ \phi  \phi}  \\
 \widehat V_{\bar{ \phi} \bar{ \phi}} &   \widehat  V_{\bar{ \phi} \phi} 
\end{bmatrix}\right]
,\quad
\rm{det}\left[ \mathcal{H}_\chi \right] = 
\rm{det}\left[ K^{\chi \bar{ \chi}}\, \begin{bmatrix}
 \widehat V_{ \chi \bar{\chi}} &  \widehat  V_{ \chi  \chi}  \\
\widehat  V_{\bar{ \chi} \bar{ \chi}} & \widehat   V_{\bar{ \chi} \chi} 
\end{bmatrix}\right],
\eea
also, due to having only one field in each sector, and $\widehat K^{\phi \bar{ \phi}}=K^{\phi \bar{ \phi}}$, $\widehat K^{\chi \bar{ \chi}}=K^{\chi \bar{ \chi}}$, the factor of the K{\"a}hler metric factors out.
We therefore can look independently for the eigenvalues of the sub-matrices   $\mathcal{H}_\phi$ and $\mathcal{H}_\chi$. The second derivatives of the  scalar potential, \eq{modified-pot}, at the local extrema are given as  follows 
\bea
\label{eq:2nderV1}
\widehat V_{\phi\phi} &=&
e^{\widetilde G}\,  \left( \partial_{\phi} \,  \partial_{\phi}   V  \right) +  \gamma\, e^{\widehat G}\, \left( K_{\phi\phi} + G_\phi  G_\phi\right)  - \Gamma^{\phi}_{\phi\phi}\,  \partial_{\phi}  \widehat V  =  e^{\widetilde G} \left( 2^{-3\alpha}\, \times 6\,   a^2 \right) , \nonumber \\ 
\widehat V_{\phi\bar\phi} &=&
e^{\widetilde G}\,   \left(  \partial_{\phi} \,  \partial_{\bar\phi}   V  \right) + \gamma\, e^{\widehat G}\,  ( K_{\phi\bar\phi} + G_\phi  G_{\bar\phi}) =  e^{\widetilde G} \left(  -2^{-3\alpha}\, \times 6\,  a^2 +  2^{-3\alpha}\, \times 9\, \gamma\,  \alpha\, a^2\right), \nonumber \\
\widehat V_{\bar\phi\bar\phi} &=&
e^{\widetilde G}\,  \left(   \partial_{\bar\phi} \,  \partial_{\bar\phi}   V \right) + \gamma\, e^{\widehat G}\,  ( K_{\bar\phi\bar\phi} + G_{\bar\phi}  G_{\bar\phi}) - \Gamma^{\bar\phi}_{\bar\phi\bar\phi}  \, \partial_{\bar\phi}  \widehat V =   e^{\widetilde G} \left( 2^{-3\alpha}\, \times 6\,   a^2 \right), 
\eea
where $e^{\widetilde G}$ is given in
 \eq{eq:G1m} and  the Christoffel symbols  in \eq{eq:Gammaor}. We note that $e^{\widehat G}\, G_\phi\, G_{\bar\phi} \neq 0$.  Finally, the eigenvalues of  the Hessian $\mathcal{H}_\phi$ at  
$\phi=\bar\phi$  are
\bea
\label{eq:eigenvals_deSR}
\lambda_1 &=& \text{e}^{\widetilde G}\, \frac{4}{3\, \alpha}  \, \left[  9\, \times 2^{- 3\, \alpha}\, a^2\, \alpha\, \gamma \right] = 3 \times \text{e}^{\widetilde G}  \times  2^{2 - 3\, \alpha}\, a^2\, \gamma , \nonumber \\
 \lambda_2 &=& \text{e}^{\widetilde G}\,  \frac{4}{3\, \alpha}\, \left[   3\, \times 2^{-3\, \alpha}\, a^2(-4+3\, \alpha \gamma) \right] =  \text{e}^{\widetilde G} \times   2^{2 -3\, \alpha}\, a^2 \left( -\frac{4}{\alpha} + 3\, \gamma \right).
\eea
Therefore, the stability condition for having a de Sitter minimum implies
\bea
3\,  \alpha \, \gamma > 4.
\eea
This shows that adding a rolling dynamics, without adding a quartic term in the K{\"a}hler metric to stabilize the imaginary part,  \eq{eq:stImqt}, also provides a way to stabilize the potential. In 
  case presented in this section, the eigenvalues of $\lambda_1$ and  $\lambda_2$, \eq{eq:eigenvals_deSR} can be both positive for suitable values of $\alpha$ and $\gamma$.

\subsubsection{Assertion of the Refined de Sitter Conjecture}
Now we check whether the modified no-scale model obeys the first criterion of RdSC,  \eq{eq:firstrdeSitter}, or not. 
We first compute the parameter $\gamma$, from \eq{eq:definitionX}
\beq 
\gamma= \left. G_{\chi}\, K^{\chi \bar\chi}\, G_{\bar\chi} \right|_{{\rm Im} \chi=0}= 
\left. \frac{1}{q |\chi|^2} \left[ q^2  |\chi|^2 + \left(\frac{p}{2}\right)^2 \ |\chi +   \overline\chi |^2  + q\ p \left(  \chi +   \overline\chi   \right)^2 \right]\right|_{{\rm Im} \chi=0}=\frac{(p+q)^2}{q},
\eeq
where 
\beq G_{\chi} = -\frac{p}{2\chi} - \frac{q}{\chi + \bar\chi}, \quad {\rm and}\quad G_{\bar\chi} = -\frac{p}{2\bar\chi} - \frac{q}{\chi + \bar\chi},
 \eeq
and $K^{\chi \bar\chi}$ is the inverse of the K{\"a}hler metric
\beq K_{\chi \bar\chi} = \frac{q}{(\chi + \bar\chi)^2}.
\eeq
The first criterion of RdSC, \eq{eq:firstrdeSitter}, in the form of \eq{eq:RdSC}, is satisfied if 
\beq
 \frac{\sqrt2(p+q)}{\sqrt q}\geq c={\cal O}(1).
\eeq
Along the real direction, we have
\beq
\label{eq:additionalG} e^{\widetilde G} = |\chi\, \bar\chi|^{-p/2}(\chi + \bar\chi)^{-q} =({\rm Re}\, \chi)^{-(p+q)} = e^{\sqrt{2\gamma}\, \chi^c},
\eeq
where $\chi^c$ is the canonically normalized rolling scalar field.  Finally, we note that if the first criterion is violated, the second criterion of the RdSC,  \eq{eq:hessian}, can be satisfied if there is an instability. In fact, the second mass eigenvalue is negative for $3\, \alpha\, \gamma < 4$. Given that 
 $\widehat{V}=12\, a^2\, \times 2^{-3\, \alpha}e^{\sqrt{2\, \gamma}\, \chi^c}$ in the real field direction, the second criterion of the RdSC, \eq{eq:hessian},  requires that
\bea 
3\, \alpha\,  \gamma \leq 4- \alpha \, c'.
\eea
Since we computed the Hessian for non-canonical fields, but kept the K{\"a}hler metric in \eq{eq:HessiandeSR} we can evaluate the second criterion, \eq{eq:hessian}, involving the fields $\phi$ and $\chi$, without the  need of using canonical fields. 

\subsubsection{Superpotential with two Terms in the Expansion of the Rolling Field \label{eq:suptwoterms}}

Now we modify the superpotential \eq{W-case1} of the minimal model to take the following form
\beq
W=(\phi^{n_+}-\phi^{n_-})(a_1\, \chi^{-p_1/2}+a_2\, \chi^{-p_2/2}).
\eeq
From \eq{eq:Gtotal} we compute 
\beq
e^{\widetilde G} = (\chi + \bar{\chi})^{-q}\, \left|a_1 \chi^{-p_1/2} + a_2\, \chi^{-p_2/2} \right|^2,
\eeq
where
\beq 
G_\chi=-\frac{p_1 + q}{2\, \chi}-\frac{(p_2-p_1)\, a_2}{2\, \chi(a_1\, \chi^{(p_2-p_1)/2} + a_2)},
\eeq
to finally get
\beq
\label{eq:gamma_twoterms}
\gamma=\frac{(p_1+q)^2}{q} + \frac{a_2\, (p_2-p_1)}{q}\left[\frac{a_2\, (p_2-p_1)}{a_1\, \chi^{(p_2-p_1)/2}+a_2}+2 (p_1+q)\right] \frac{1}{a_1\, \chi^{(p_2-p_1)/2} + a_2}.
\eeq
From \eq{eq:gamma_twoterms} we observe that as far as $\sqrt{2\gamma}\geq c$ the RdSC, in the form of  \eq{eq:RdSC},  can be satisfied.

\section{Non-minimal No-Scale  Models \label{sec:nonminimal}}
\paragraph*{2+1 Model} 
Next, we consider a no-scale supergravity with dS vacuum generated by two chiral superfields $\phi_i$ and add one rolling superfield $\chi$,  we refer to this construction as the \emph{2+1 Model}.  The K{\"a}hler and the superpotential are as follows
\bea
K &=& -3\, \alpha_1 \ln (\phi_1 + \bar{\phi}_1) - 3\, \alpha_2\, \ln (\phi_2 + \bar{\phi}_2) - q\, \ln (\chi + \bar{\chi}),\\
\label{W-case2p1}W&=& a \left(\phi_1^{n_{1+}}\, \phi_2^{n_{2+}} - \phi_1^{n_{1-}}\, \phi_2^{n_{2-}}  \right) \, \chi^{-p/2}
,\eea
where $\alpha_i,q>0, p<0$, $a$ is an arbitrary constant and $n_{i\pm}$ are given in \eq{eq:ni_pm}.  From \eq{eq:min_Adterm} and demanding $n_{i+}\neq n_{i-}$, we have
\beq 
\label{sol-case2} 
\phi_1^{r_1}\phi_2^{r_2}=\bar\phi_1^{r_1}\bar\phi_2^{r_2}=1,
\eeq
where $r_1,r_2$ are arbitrary constants satisfying $r_1^2+r_2^2=1$. We note 
 that we cannot have both 
$\partial_1\, W$ and $\partial_2\, W$ 
simultaneously vanishing  as supersymmetry is spontaneously broken and the scalar potential is lifted. However,  one of them can vanish for special case of $r_1=0$ and $r_2=1$, or vice versa. Moreover, we observe that the above solution \eq{sol-case2} satisfies \eq{eq:V0min}.  Similar to the minimal model, the Hessian matrix can be brought to a block diagonal form, \eq{eq:blockdiagonalH},
where now
\bea
\label{eq:4by4Hessian}
{\mathcal{H}}_\phi= \left[
\begin{array}{cccc}
 K^{1\bar 1}\, \widehat V_{1 \bar 1}  &  K^{1\bar 1}\,  g_{1\bar 1}  \, \widehat V_{1 1} &K^{1\bar 1}\, \widehat V_{1 \bar 2}  &   K^{1\bar 1}\,  g_{1\bar 1} \widehat V_{1  2} \\
  K^{1\bar 1}\, g^{1\bar 1}  \widehat V_{\bar 1 \bar1}  &   K^{1\bar 1}\,  \widehat V_{\bar 1 1}  &   K^{1\bar 1}\, g^{1\bar 1} \, \widehat V_{\bar 1 \bar 2}  &   K^{1\bar 1}\, \widehat V_{\bar 1  2} \\
 K^{2\bar 2}\, g^{2 \bar2}\,  \widehat V_{2 \bar 1}  &  K^{2\bar 2}\, g_{2\bar 2} \, \widehat V_{2 1}  & K^{2 \bar 2} \widehat V_{2 \bar 2}  &   K^{2\bar 2}\, g_{2\bar 2} \, \widehat V_{2  2} \\
 K^{ 2 \bar 2}\,  g^{2\bar 2} \, \widehat V_{\bar 2 \bar 1}  & K^{2 \bar 2}\, \widehat V_{\bar 2 1} &   K^{2\bar 2}\, \, g^{2\bar 2}  \widehat V_{\bar 2 \bar 2}  &  K^{2\bar 2}\, \widehat V_{\bar 2  2}
\end{array}
\right],
\eea
and  ${\mathcal{H}}_\chi$ is as in \eq{eq:Hessians}, therefore the problem reduces to analyzing the $4\times 4$  ${\mathcal{H}}_1$ matrix of \eq{eq:4by4Hessian}.  Analogously to \eq{eq:2nderV1} we need to compute the 
second order derivatives of $\widehat V$ with respect to $\phi_1$ and $\phi_2$:
$\widehat V_{\phi_i \phi_j}=\partial_i\,\partial_j\, {\rm{e}}^{\widetilde G} \left(V + \gamma\, e^G \right)=  $
$ {\rm{e}}^{\widetilde G}\, \partial_i\,\partial_j\, \left(V + \gamma\, e^G \right)$. These derivatives are given in the Appendix \ref{app:SecDer2fields}. We find that the eigenvalues are  given as follows
\bea
&& \lambda_1=0, \nonumber\\
{e^{-\tilde G}} \hspace*{-0.4 cm} &&  \lambda_2= \frac{12\, \times 2^{-3\, (\alpha_1 + \alpha_2)}a^2(3\, r_2^2\, \alpha_1+3\, r_1^2\, \alpha_2+s^2 \,\alpha_1 \, \alpha_2 (\gamma-3))(r_2^2\alpha_1 +r_1^2\alpha_2)}{s^4\, \alpha_1^2\, \alpha_2^2},
\eea
where $s$ has been defined in \eq{eq:ni_pm} , 
which requires 
\bea
 3\, r_2^2\, \alpha_1 + 3\, r_1^2 \,\alpha_2+s^2\, \alpha_1\, \alpha_2 (\gamma-3) > 0,
\eea
while the other two eigenvalues are of the form
\bea
{e^{-\tilde G}} \,  \lambda_\pm= A \pm \sqrt{B},
\eea
where $A$ and $B$ are independent of $\phi_1$ and $\phi_2$. In  \Figref{fig:2p1model} we plot the possible values of $\alpha_1$ and $\alpha_2$ for the special case of $\gamma=2$ and $c'=1$ with $r_1=r_2=\frac{1}{\sqrt{2}}$ 
that are  compatible either with the RdSC or the stability condition.  

\begin{figure}
\label{fig:2p1model}
\begin{center}
\includegraphics[width=9cm]{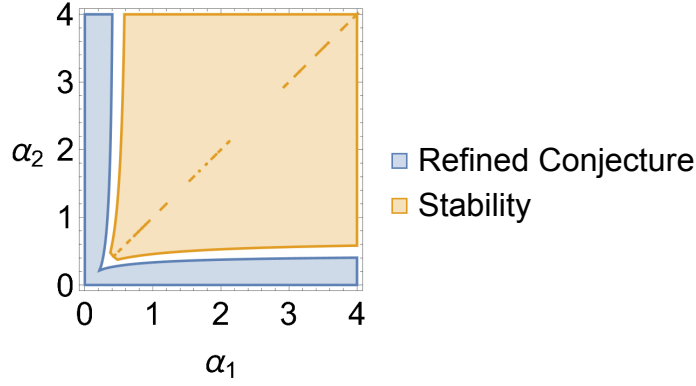}
\end{center}
\caption{Parameter space of $\alpha_1$ and $\alpha_2$, for the $2+1$ model  for the special case of $\gamma=2$ and $c'=1$ with $r_1=r_2=\frac{1}{\sqrt{2}}$, that can be either compatible with the RdSC or the stability condition. }
\end{figure}

\paragraph*{N+1 Model}
We generalize to the case of N de Sitter case plus a rolling field with the following  general K{\"a}hler  potential and superpotential 
\bea
K &=& -3 \sum_{i=1}^N\alpha_i\ln (\phi_i + \bar{\phi}_i)- q \ln (\chi + \bar{\chi}) = K - q \ln (\chi + \bar{\chi}),\\
\label{W-case3}W&=& a\, \left(\prod_{i=1}^N\phi_i^{n_{i+}} - \prod_{i=1}^N\phi_i^{n_{i-}}   \right)\, \chi^{-p/2} = W\chi^{-p/2}.
\eea
In this case, local supersymmetry breaking vacua exist when for some fields  \eq{eq:min_Adterm} is satisfied (leading to $W=\overline W=0$), which can be put in the form 
\beq 
\label{sol-case3} \prod_{i=1}^N\phi_i^{r_i}=\prod_{i=1}^N\bar\phi_i^{r_i}=1,
\eeq
where $r_1,r_2$ are arbitrary constants satisfying $\sum_{i=1}^Nr_i^2=1$. We note that we cannot have  $\partial_i\, W$ simultaneously vanishing for all $i$. However, some of them can vanish for special values of $r_i$.  As with the previous examples this solution automatically satisfies $\partial_i\, V=0$, \eq{eq:V0min}, which requires $\phi = \bar\phi$. The part of the Hessian  regarding only $\phi_i$ fields does not mix with that of $\chi_i$, hence as in the previous case the total Hessian is block diagonal, \eq{eq:blockdiagonalH},
where
\bea
{\mathcal{H}}_\phi= \left[
\begin{array}{ccccc}
\widehat V_{\phi_1\bar \phi_1}   &  \widehat V_{\phi_1\bar \phi_2}  & \hdots &  \hdots &  \widehat V_{\phi_1 \phi_N}  \\
 \widehat V_{\phi_2 \bar \phi_1}   &  \widehat V_{\phi_2 \bar \phi_2}   & \hdots & \hdots &\widehat V_{\phi_2 \phi_N} \\
  \vdots  &  \vdots  & \vdots & \vdots &\vdots \\
   \widehat V_{\bar\phi_N \phi_1}  &  \hdots   &  \hdots & \hdots &  \widehat V_{\bar\phi_N \phi_N} 
\end{array}
\right],
\eea
and  ${\mathcal{H}}_\chi$ is as in \eq{eq:Hessians}, therefore the problem reduces to analysing the $2\, N \times 2\, N$  ${\mathcal{H}}_\phi$ matrix, which in principle is not block diagonal itself.

\section{Application to Quintessence Models \label{sec:Quintessence}}

Current data indicates that the universe is dominated by dark energy, therefore if \eq{eq:firstrdeSitter} is 
satisfied, this implies that this cannot be the result of a positive cosmological constant nor can be described by a state at the minimum of a potential with positive energy density. In such a case we 
 can invoke Quintessence models, which are described by rolling fields.  It is known that exponential potentials of the form  $V=V_0 \, e^{\lambda\,  \chi}$ \cite{Agrawal:2018own}  and 
\bea
\label{eq:scalingsols}
V(\chi)=V_1 \, e^{\lambda_1 \, \chi}\,  +\, V_2 \, e^{\lambda_1\, \chi },
\eea
with varying $\lambda_i$ parameters  \cite{Chiba:2012cb} that can fit well the constraints based on  Supernovae type Ia (SNeIa), Cosmic Microwave Background (CMB) and Baryon Acoustic Oscillations  (BAO) data  (see \cite{Akrami_2018}  for an alternative conclusion).  For the case of one rolling field $\chi$, and hence only one term in \eq{eq:scalingsols}, $V(\chi)=V_1 \, e^{\lambda \,\chi}$,   the variable $\lambda$ is constrained to be $\lambda=c$, which is the constant appearing in \eq{eq:firstrdeSitter}, and must satisfy $\lambda \lesssim 0.6$. In \cite{Ferrara:2019tmu} the mechanism that we use in this work to satisfy the RdSC was applied to uplift a vacuum to a de Sitter vaccum that could be in agreement with the RdSC but it was concluded that the breaking of supersymmetry  needed for the de Sitter uplift cannot be only caused by the Quintessence field $\chi$ due to the requirement $\gamma=c>3$.  In the models we present in \cref{sec:noscale_rollingdyn} we do not have such a requirement and hence even our minimal model with one rolling field plus one de Sitter field could be used as a Quintessence model.  In particular our result for the minimal model for the constraint on $\gamma$ is compatible with a value of $0.6$.

The potential with two terms,  \eq{eq:scalingsols}, is in fact a kind of freezing model associated with scaling solutions \cite{Copeland:1997et}, where the field equation of state scales as that of the background fluid during most of the matter era. Here the constants $\lambda_i$ are constrained to $\lambda_1 \gg 1$ and  $\lambda_2 \lesssim 1$ \cite{Chiba:2012cb}. In the early matter era, the potential is approximated by the first term in \eq{eq:scalingsols} while at late times the second term.  In \cite{Agrawal:2018own} it was shown that the potential of \eq{eq:scalingsols} with $\lambda \approx \lambda_1 >> \sqrt{3} $ in the early universe and then switches to $\lambda \approx \lambda_2 =c = 0.6 $ at some recent point in the past. Together these two stages approxiate the boundary trajectory that the dark energy equation of state, $\omega(a)=\omega_0 + \omega_a (1-a)$, where $a$ is the scale factor normalized as $a=1$ today, needs to satisfy in order to be in agreement with  SNeIa, CMB and BAO and the RdSC, \eq{eq:firstrdeSitter}, as shown in \cite{Agrawal:2018own}. The No-Scale models that we study in \cref{eq:suptwoterms} can easily accommodate these requirements.  We are aware about the difficulties in constructing Quintessence models in supergravity \cite{Brax:2006np}, where  either it is difficult to evade gravitational tests or having the Quintessence/supegravity models behave like pure cosmological constants. Nevertheless, our result is encouraging from the viewpoint of making compatible a supergravity theory with the Refined de Sitter conjecture {{\footnote{There is a  current debate on the value of the Hubble constant,   \cite{Riess:2018byc,Riess:2020fzl}  which may lead to a revolution in cosmology. Solutions for this tension may even discard Quintessence models, \cite{Krishnan:2021dyb,Banerjee_2021}. Nevertheless, one can obviously not rule out  Quintessence models on this basis, as the model realized in nature resolving the tension has not been established yet. In addition, there may not even be a need for new physics as potential reconsiderations of instrument calibration or data analysis may hold a key in resolving the Hubble tension. For example, Cepheid calibration reconsiderations can settle on a value in agreement with Standard Cosmology \cite{mortsell2021hubble}.}.}}

\section{Conclusions   \label{sec:Conclusion}}

In this paper, we used the Refined de Sitter Conjecture (RdSC) and the Trans-Planckian Censorship  Conjecture (TCC) to constrain the parameter space of some no-scale supergravity models. Specifically, we have studied a toy model of one field with no-scale K{\"a}hler potential and a superpotential constructed from two Minkowski endpoints, which yields a de Sitter scalar potential. 
It is well known that for a particular choice of parameters, this
model is unstable along the imaginary direction, that is the second
derivative of the scalar potential is negative. But this is exactly
the kind of model that the RdSC allows and the TCC can constrain. To
check the compatibility with the RdSC and the TCC we first added a
quartic term to the original K{\"a}hler potential to stabilize the
real part of the field, allowing us to work with one canonical field
that is proportional to the imaginary direction. Using this, we have shown that for some choices of parameters the theory can be compatible with the RdSC and the TCC.
This analysis can be extended to the case of multi-field theories as
well as to cases with meta-stable de Sitter vacua, where the
imaginary direction can also be modified via a quartic term in the
K{\"a}hler potential. We think this kind of analysis is important as no-scale models with Minkowski/Anti-de Sitter vacua generically appear as low energy limits of string compactifications, but de Sitter vacua \emph{``have not yet been rigorously shown to be realized in string theory'' }\cite{Obied:2018sgi}.

To construct the second class of models we presented in this paper, for which an effective dS background is obtained from a rolling
potential, we modified the K{\"a}hler and superpotential  of the
simplest no-scale model we considered by adding  the terms
corresponding to the rolling fields.  We found that this modification
alleviates the instability and flatness along the imaginary and the
real directions without the need for quartic terms. The existence of the rolling direction provides an escape from the swamp as the first criterion of the RdSC can be satisfied for suitable values of parameters. Interestingly, we found that the height of the potential is controlled by the no-scale parameters while the rolling parameters make the model compatible with the RdSC. These models could be used to construct viable cosmological models to accommodate the accelerating expansion of the late Universe.

\section*{Acknowledgements}

L. V.-S. acknowledges the support from the ``Fundamental Research Program"  of the Korea Institute for Advanced Study and the warm hospitality and stimulating environment.  M. T. is supported by the research deputy of Sharif University of Technology.  We are grateful to K. Kaneta, J. Kersten and K. Olive for helpful discussions and J. Kersten and F. Borzumati for comments on the manuscript. 

\appendix 
\section{\textit{Canonical Normalization} \label{app:cannorm}}
To evaluate the second criterion of the RdSC,  \eq{eq:hessian}, we split the complex field $\phi$ into real and imaginary components $\phi= x + i y$, to have
\bea
{\mathcal{L}}= - K_{\phi \bar\phi}\, \partial_\mu \phi \, \partial^\mu \bar\phi=
- \, \frac{3 \alpha}{4\, [{\rm{Re}}[\phi]]^2} \,\left[ (\partial_\mu x)(\partial^\mu x) + (\partial_\mu y)(\partial^\mu y) \right]. 
\eea
With the K{\"a}hler metric of   \eq{eq:Kahler_quartic_realterm} we can fix the real part to b, such that 
\bea
\langle {\rm{Re}}[\phi]  \rangle= b.
\eea
Then we identify the canonical kinetic term of the fields $\chi$ and $\sigma$:
\bea
{\mathcal{L}}= - \left( \frac{1}{\sqrt{2}} \partial_\mu \chi \right)^2  - \left( \frac{1}{\sqrt{2}} \partial_\mu \sigma \right)^2,
\eea
we would have
\bea
\frac{1}{\sqrt{2}}\, \partial_\mu \chi = \sqrt{ \frac{3 \alpha}{4\, [{\rm{Re}}[\phi]]^2}   } \, \partial_\mu x \, ,
\eea
\bea
\label{eq:transfO}
\frac{1}{\sqrt{2}}\, \partial_\mu \sigma =  \sqrt{ \frac{3 \alpha}{4\,}   } \,  \frac{\partial_\mu y}{ b}\,
\rightarrow \,  \sigma= \sqrt{\frac{3 \alpha}{2}} \, \frac{y}{b}.
\eea
But since the real part is fixed and acquires a vev, then our effective theory will consist in only one field, $\sigma$, with canonical kinetic term ${\mathcal{L}} =-\frac{1}{2}\, \partial_\mu \sigma \partial^\mu \sigma$. 
\section{\textit{Details of Stability Conditions} \label{app:SecDer2fields}}
For the  \emph{2+1 Model} the second derivatives of $\widehat V$ are as follows 

\bea
\frac{1}{e^{\widetilde G}} K^1_1\widehat V^{1}_{\; \; 1} &=& \frac{1}{e^{\widetilde G}} K^1_1 \widehat V_1^{\; 1}  = \frac{4\, \times 2^{-3\, (\alpha_1 + \alpha_2)}a^2(3r_2^2 \alpha_1^2-2r_1^2 \alpha_2+3\, \gamma \alpha_1 \alpha_2 r_1^2)}{s^2 \alpha_1^2 \alpha_2},\nonumber\\
\frac{1}{e^{\widetilde G}} K^1_1\widehat V^{1 1}&=& \frac{1}{e^{\widetilde G}} K^1_1\widehat V_{1 1} =\frac{4\, \times 2^{-3\, (\alpha_1 + \alpha_2)}a^2(s^2\, (2\, r_1^2 \alpha_2 -9\, r_1^2 \alpha_1 \alpha_2 -3r_2^2 \alpha_1^2)+9\, r_1^2(r_2^2 \alpha_1+r_1^2 \alpha_2))}{s^4 \alpha_1^2 \alpha_2}, \nonumber\\
 \frac{1}{e^{\widetilde G}} K^1_1 \widehat V^{1}_{\; \; 2} &=&  \frac{1}{e^{\widetilde G}}K^1_1\widehat V_1^{\; 2} =\frac{12\, \, \times 2^{-3\, (\alpha_1 + \alpha_2)}a^2 r_1 r_2(\gamma-1)\phi_1}{s^2\, \alpha_1 \phi_2},\nonumber\\
\frac{1}{e^{\widetilde G}}K^1_1\widehat V^{1 2} &=& \frac{1}{e^{\widetilde G}} K^1_1\widehat V_{1 2} =\frac{12\, \, \times 2^{-3\, (\alpha_1 + \alpha_2)}a^2\,  r_1 r_2(3r_1^2 \alpha_2+3r_2^2 \alpha_1-2\, s^2 \alpha_1 \alpha_2)\phi_1}{s^4 \alpha_1^2 \alpha_2 \phi_2},
\nonumber\\
 \frac{1}{e^{\widetilde G}} K^2_2\widehat V^{2}_{\; \; 1} &=&  \frac{1}{e^{\widetilde G}}K^2_2 \widehat V_2^{\; 1} =\frac{12\, \, \times 2^{-3\, (\alpha_1 + \alpha_2)}a^2 r_1 r_2(\gamma-1)\phi_2}{s^2\, \alpha_2 \phi_1},\nonumber\\
 \frac{1}{e^{\widetilde G}} K^2_2\widehat V^{2 1} &=& \frac{1}{e^{\widetilde G}} K^2_2\widehat V_{2 1} = \frac{12\, \, \times 2^{-3\, (\alpha_1 + \alpha_2)}a^2\,  r_1 r_2(3r_1^2 \alpha_2+3r_2^2 \alpha_1-2\, s^2 \alpha_1 \alpha_2)\phi_2}{s^4 \alpha_2^2 \alpha_1 \phi_1},
\nonumber\\
 \frac{1}{e^{\widetilde G}} K^2_2\widehat V^{2 2} &=&  \frac{1}{e^{\widetilde G}}K^2_2\widehat V_{2 2} = \frac{4\, \times 2^{-3\, (\alpha_1 + \alpha_2)}a^2(s^2\, (2\, r_2^2 \alpha_1 -9\, r_2^2 \alpha_2 \alpha_1 -3r_1^2 \alpha_2^2)+9\, r_2^2(r_1^2 \alpha_2+r_2^2 \alpha_1))}{s^4 \alpha_2^2 \alpha_1}, \nonumber \\
 \frac{1}{e^{\widetilde G}} K^2_2 \widehat V^{2}_{\; \; 2} &=&  \frac{1}{e^{\widetilde G}}K^2_2 \widehat V_2^{\; 2} = \frac{4\, \times 2^{-3\, (\alpha_1 + \alpha_2)}a^2(3r_1^2 \alpha_2^2-2r_2^2 \alpha_1+3\, \gamma \alpha_2 \alpha_1 r_2^2)}{s^2 \alpha_2^2 \alpha_1}.
\eea



\begin{thebibliography}{10}
\providecommand{\url}[1]{#1}
\csname url@samestyle\endcsname
\providecommand{\newblock}{\relax}
\providecommand{\bibinfo}[2]{#2}
\providecommand{\BIBentrySTDinterwordspacing}{\spaceskip=0pt\relax}
\providecommand{\BIBentryALTinterwordstretchfactor}{4}
\providecommand{\BIBentryALTinterwordspacing}{\spaceskip=\fontdimen2\font plus
\BIBentryALTinterwordstretchfactor\fontdimen3\font minus
  \fontdimen4\font\relax}
\providecommand{\BIBforeignlanguage}[2]{{%
\expandafter\ifx\csname l@#1\endcsname\relax
\typeout{** WARNING: IEEEtran.bst: No hyphenation pattern has been}%
\typeout{** loaded for the language `#1'. Using the pattern for}%
\typeout{** the default language instead.}%
\else
\language=\csname l@#1\endcsname
\fi
#2}}
\providecommand{\BIBdecl}{\relax}
\BIBdecl
\renewcommand{\BIBentryALTinterwordstretchfactor}{4}

\bibitem{Kachru:2003aw}
\BIBentryALTinterwordspacing
S.  Kachru, R.  Kallosh, A.~D.  Linde, and S.~P.  Trivedi, ``{De Sitter vacua
  in string theory},'' \emph{Phys. Rev. D}, vol.~68, p. 046005, 2003.
  \url{https://arxiv.org/abs/hep-th/0301240}
\BIBentrySTDinterwordspacing

\bibitem{Balasubramanian:2005zx}
\BIBentryALTinterwordspacing
V.  Balasubramanian, P.  Berglund, J.~P.  Conlon, and F.  Quevedo,
  ``{Systematics of moduli stabilisation in Calabi-Yau flux
  compactifications},'' \emph{JHEP}, vol.~03, p. 007, 2005.
  \url{https://arxiv.org/abs/hep-th/0502058}
\BIBentrySTDinterwordspacing

\bibitem{Westphal:2006tn}
\BIBentryALTinterwordspacing
A.  Westphal, ``{de Sitter string vacua from Kahler uplifting},'' \emph{JHEP},
  vol.~03, p. 102, 2007.  \url{https://arxiv.org/abs/hep-th/0611332}
\BIBentrySTDinterwordspacing

\bibitem{Rummel:2011cd}
\BIBentryALTinterwordspacing
M.  Rummel and A.  Westphal, ``{A sufficient condition for de Sitter vacua in
  type IIB string theory},'' \emph{JHEP}, vol.~01, p. 020, 2012.
  \url{https://arxiv.org/abs/1107.2115}
\BIBentrySTDinterwordspacing

\bibitem{Blaback:2013fca}
\BIBentryALTinterwordspacing
J.  Bl\r{a}b\"ack, U.  Danielsson, and G.  Dibitetto, ``{Accelerated Universes
  from type IIA Compactifications},'' \emph{JCAP}, vol.~03, p. 003, 2014.
  \url{https://arxiv.org/abs/1310.8300}
\BIBentrySTDinterwordspacing

\bibitem{Cicoli:2013cha}
\BIBentryALTinterwordspacing
M.  Cicoli, D.  Klevers, S.  Krippendorf, C.  Mayrhofer, F.  Quevedo, and R.
  Valandro, ``{Explicit de Sitter Flux Vacua for Global String Models with
  Chiral Matter},'' \emph{JHEP}, vol.~05, p. 001, 2014.
  \url{https://arxiv.org/abs/1312.0014}
\BIBentrySTDinterwordspacing

\bibitem{Cicoli:2015ylx}
\BIBentryALTinterwordspacing
M.  Cicoli, F.  Quevedo, and R.  Valandro, ``{De Sitter from T-branes},''
  \emph{JHEP}, vol.~03, p. 141, 2016.  \url{https://arxiv.org/abs/1512.04558}
\BIBentrySTDinterwordspacing

\bibitem{Heckman:2018mxl}
\BIBentryALTinterwordspacing
J.~J.  Heckman, C.  Lawrie, L.  Lin, and G.  Zoccarato, ``{F-theory and Dark
  Energy},'' \emph{Fortsch. Phys.}, vol.~67, no.~10, p. 1900057, 2019.
  \url{https://arxiv.org/abs/1811.01959}
\BIBentrySTDinterwordspacing

\bibitem{Heckman:2019dsj}
\BIBentryALTinterwordspacing
J.~J.  Heckman, C.  Lawrie, L.  Lin, J.  Sakstein, and G.  Zoccarato,
  ``{Pixelated Dark Energy},'' \emph{Fortsch. Phys.}, vol.~67, no.~11, p.
  1900071, 2019.  \url{https://arxiv.org/abs/1901.10489}
\BIBentrySTDinterwordspacing

\bibitem{Riess:1998cb}
\BIBentryALTinterwordspacing
A.~G.  Riess \emph{et~al.}, ``{Observational evidence from supernovae for an
  accelerating universe and a cosmological constant},'' \emph{Astron. J.}, vol.
  116, pp. 1009--1038, 1998.  \url{https://arxiv.org/abs/astro-ph/9805201}
\BIBentrySTDinterwordspacing

\bibitem{Riess:1998dv}
\BIBentryALTinterwordspacing
------, ``{BV RI light curves for 22 type Ia supernovae},'' \emph{Astron. J.},
  vol. 117, pp. 707--724, 1999.  \url{https://arxiv.org/abs/astro-ph/9810291}
\BIBentrySTDinterwordspacing

\bibitem{Akrami:2018odb}
\BIBentryALTinterwordspacing
Y.  Akrami \emph{et~al.}, ``{Planck 2018 results. X. Constraints on
  inflation},'' \emph{Astron. Astrophys.}, vol. 641, p. A10, 2020.
  \url{https://arxiv.org/abs/1807.06211}
\BIBentrySTDinterwordspacing

\bibitem{Maldacena:2000mw}
\BIBentryALTinterwordspacing
J.~M.  Maldacena and C.  Nunez, ``{Supergravity description of field theories
  on curved manifolds and a no go theorem},'' \emph{Int. J. Mod. Phys. A},
  vol.~16, pp. 822--855, 2001.  \url{https://arxiv.org/abs/hep-th/0007018}
\BIBentrySTDinterwordspacing

\bibitem{Hertzberg:2007wc}
\BIBentryALTinterwordspacing
M.~P.  Hertzberg, S.  Kachru, W.  Taylor, and M.  Tegmark, ``{Inflationary
  Constraints on Type IIA String Theory},'' \emph{JHEP}, vol.~12, p. 095, 2007.
   \url{https://arxiv.org/abs/0711.2512}
\BIBentrySTDinterwordspacing

\bibitem{Covi:2008ea}
\BIBentryALTinterwordspacing
L.  Covi, M.  Gomez-Reino, C.  Gross, J.  Louis, G.~A.  Palma, and C.~A.
  Scrucca, ``{de Sitter vacua in no-scale supergravities and Calabi-Yau string
  models},'' \emph{JHEP}, vol.~06, p. 057, 2008.
  \url{https://arxiv.org/abs/0804.1073}
\BIBentrySTDinterwordspacing

\bibitem{deCarlos:2009fq}
\BIBentryALTinterwordspacing
B.  de~Carlos, A.  Guarino, and J.~M.  Moreno, ``{Flux moduli stabilisation,
  Supergravity algebras and no-go theorems},'' \emph{JHEP}, vol.~01, p. 012,
  2010.  \url{https://arxiv.org/abs/hep-th/0907.5580}
\BIBentrySTDinterwordspacing

\bibitem{Wrase:2010ew}
\BIBentryALTinterwordspacing
T.  Wrase and M.  Zagermann, ``{On Classical de Sitter Vacua in String
  Theory},'' \emph{Fortsch. Phys.}, vol.~58, pp. 906--910, 2010.
  \url{https://arxiv.org/abs/1003.0029}
\BIBentrySTDinterwordspacing

\bibitem{Shiu:2011zt}
\BIBentryALTinterwordspacing
G.  Shiu and Y.  Sumitomo, ``{Stability Constraints on Classical de Sitter
  Vacua},'' \emph{JHEP}, vol.~09, p. 052, 2011.
  \url{https://arxiv.org/abs/1107.2925}
\BIBentrySTDinterwordspacing

\bibitem{Green:2011cn}
\BIBentryALTinterwordspacing
S.~R.  Green, E.~J.  Martinec, C.  Quigley, and S.  Sethi, ``{Constraints on
  String Cosmology},'' \emph{Class. Quant. Grav.}, vol.~29, p. 075006, 2012.
  \url{https://arxiv.org/abs/1110.0545}
\BIBentrySTDinterwordspacing

\bibitem{Dasgupta:2014pma}
\BIBentryALTinterwordspacing
K.  Dasgupta, R.  Gwyn, E.  McDonough, M.  Mia, and R.  Tatar, ``{de Sitter
  Vacua in Type IIB String Theory: Classical Solutions and Quantum
  Corrections},'' \emph{JHEP}, vol.~07, p. 054, 2014.
  \url{https://arxiv.org/abs/1402.5112}
\BIBentrySTDinterwordspacing

\bibitem{Kutasov:2015eba}
\BIBentryALTinterwordspacing
D.  Kutasov, T.  Maxfield, I.  Melnikov, and S.  Sethi, ``{Constraining de
  Sitter Space in String Theory},'' \emph{Phys. Rev. Lett.}, vol. 115, no.~7,
  p. 071305, 2015.  \url{https://arxiv.org/abs/1504.00056}
\BIBentrySTDinterwordspacing

\bibitem{Quigley:2015jia}
\BIBentryALTinterwordspacing
C.  Quigley, ``{Gaugino Condensation and the Cosmological Constant},''
  \emph{JHEP}, vol.~06, p. 104, 2015.  \url{https://arxiv.org/abs/1504.00652}
\BIBentrySTDinterwordspacing

\bibitem{Junghans:2016uvg}
\BIBentryALTinterwordspacing
D.  Junghans, ``{Tachyons in Classical de Sitter Vacua},'' \emph{JHEP},
  vol.~06, p. 132, 2016.  \url{https://arxiv.org/abs/1603.08939}
\BIBentrySTDinterwordspacing

\bibitem{Andriot:2016xvq}
\BIBentryALTinterwordspacing
D.  Andriot and J.  Bl\r{a}b\"ack, ``{Refining the boundaries of the classical
  de Sitter landscape},'' \emph{JHEP}, vol.~03, p. 102, 2017, [Erratum: JHEP
  03, 083 (2018)].  \url{https://arxiv.org/abs/1609.00385}
\BIBentrySTDinterwordspacing

\bibitem{Junghans:2016abx}
\BIBentryALTinterwordspacing
D.  Junghans and M.  Zagermann, ``{A Universal Tachyon in Nearly No-scale de
  Sitter Compactifications},'' \emph{JHEP}, vol.~07, p. 078, 2018.
  \url{https://arxiv.org/abs/1612.06847}
\BIBentrySTDinterwordspacing

\bibitem{Moritz:2017xto}
\BIBentryALTinterwordspacing
J.  Moritz, A.  Retolaza, and A.  Westphal, ``{Toward de Sitter space from ten
  dimensions},'' \emph{Phys. Rev. D}, vol.~97, no.~4, p. 046010, 2018.
  \url{https://arxiv.org/abs/1707.08678}
\BIBentrySTDinterwordspacing

\bibitem{Sethi:2017phn}
\BIBentryALTinterwordspacing
S.  Sethi, ``{Supersymmetry Breaking by Fluxes},'' \emph{JHEP}, vol.~10, p.
  022, 2018.  \url{https://arxiv.org/abs/1709.03554}
\BIBentrySTDinterwordspacing

\bibitem{Danielsson:2018ztv}
\BIBentryALTinterwordspacing
U.~H.  Danielsson and T.  Van~Riet, ``{What if string theory has no de Sitter
  vacua?}'' \emph{Int. J. Mod. Phys. D}, vol.~27, no.~12, p. 1830007, 2018.
  \url{https://arxiv.org/abs/1804.01120}
\BIBentrySTDinterwordspacing

\bibitem{Ooguri:2018wrx}
\BIBentryALTinterwordspacing
H.  Ooguri, E.  Palti, G.  Shiu, and C.  Vafa, ``{Distance and de Sitter
  Conjectures on the Swampland},'' \emph{Phys. Lett. B}, vol. 788, pp.
  180--184, 2019.  \url{https://arxiv.org/abs/1810.05506}
\BIBentrySTDinterwordspacing

\bibitem{Obied:2018sgi}
\BIBentryALTinterwordspacing
G.  Obied, H.  Ooguri, L.  Spodyneiko, and C.  Vafa, ``{De Sitter Space and the
  Swampland},'' 6 2018.  \url{https://arxiv.org/abs/1806.08362}
\BIBentrySTDinterwordspacing

\bibitem{Garg:2018reu}
\BIBentryALTinterwordspacing
S.~K.  Garg and C.  Krishnan, ``{Bounds on Slow Roll and the de Sitter
  Swampland},'' \emph{JHEP}, vol.~11, p. 075, 2019.
  \url{https://arxiv.org/abs/1807.05193}
\BIBentrySTDinterwordspacing

\bibitem{Bedroya:2019snp}
\BIBentryALTinterwordspacing
A.  Bedroya and C.  Vafa, ``{Trans-Planckian Censorship and the Swampland},''
  \emph{JHEP}, vol.~09, p. 123, 2020.  \url{https://arxiv.org/abs/1909.11063}
\BIBentrySTDinterwordspacing

\bibitem{Andriot_2020}
\BIBentryALTinterwordspacing
D.  Andriot, N.  Cribiori, and D.  Erkinger, ``The web of swampland conjectures
  and the tcc bound,'' \emph{Journal of High Energy Physics}, vol. 2020, no.~7,
  Jul 2020.  \url{http://dx.doi.org/10.1007/JHEP07(2020)162}
\BIBentrySTDinterwordspacing

\bibitem{bedroya2020sitter}
A.  Bedroya, ``de sitter complementarity, tcc, and the swampland,'' 2020.

\bibitem{Aalsma_2020}
\BIBentryALTinterwordspacing
L.  Aalsma and G.  Shiu, ``Chaos and complementarity in de sitter space,''
  \emph{Journal of High Energy Physics}, vol. 2020, no.~5, May 2020.
  \url{http://dx.doi.org/10.1007/JHEP05(2020)152}
\BIBentrySTDinterwordspacing

\bibitem{aalsma2021shocks}
L.  Aalsma, A.  Cole, E.  Morvan, J.~P.  van~der Schaar, and G.  Shiu, ``Shocks
  and information exchange in de sitter space,'' 2021.

\bibitem{Palti:2019pca}
\BIBentryALTinterwordspacing
E.  Palti, ``{The Swampland: Introduction and Review},'' \emph{Fortsch. Phys.},
  vol.~67, no.~6, p. 1900037, 2019.  \url{https://arxiv.org/abs/1903.06239}
\BIBentrySTDinterwordspacing

\bibitem{Ellis:1983ei}
\BIBentryALTinterwordspacing
J.~R.  Ellis, C.  Kounnas, and D.~V.  Nanopoulos, ``{Phenomenological SU(1,1)
  Supergravity},'' \emph{Nucl. Phys. B}, vol. 241, pp. 406--428, 1984.
  \url{https://doi.org/10.1016/0550-3213(84)90054-3}
\BIBentrySTDinterwordspacing

\bibitem{Roest:2015qya}
\BIBentryALTinterwordspacing
D.  Roest and M.  Scalisi, ``{Cosmological attractors from
  \ensuremath{\alpha}-scale supergravity},'' \emph{Phys. Rev. D}, vol.~92, p.
  043525, 2015.  \url{https://arxiv.org/abs/1503.07909}
\BIBentrySTDinterwordspacing

\bibitem{Ellis:2018xdr}
\BIBentryALTinterwordspacing
J.  Ellis, B.  Nagaraj, D.~V.  Nanopoulos, and K.~A.  Olive, ``{De Sitter Vacua
  in No-Scale Supergravity},'' \emph{JHEP}, vol.~11, p. 110, 2018.
  \url{https://arxiv.org/abs/1809.10114}
\BIBentrySTDinterwordspacing

\bibitem{Ellis:2019hps}
\BIBentryALTinterwordspacing
J.  Ellis, B.  Nagaraj, D.~V.  Nanopoulos, K.~A.  Olive, and S.  Verner,
  ``{From Minkowski to de Sitter in Multifield No-Scale Models},'' \emph{JHEP},
  vol.~10, p. 161, 2019.  \url{https://arxiv.org/abs/1907.09123}
\BIBentrySTDinterwordspacing

\bibitem{Ellis:2019dtx}
\BIBentryALTinterwordspacing
J.  Ellis, D.~V.  Nanopoulos, K.~A.  Olive, and S.  Verner, ``{Unified no-scale
  model of modulus fixing, inflation, supersymmetry breaking, and dark
  energy},'' \emph{Phys. Rev. D}, vol. 100, no.~2, p. 025009, 2019.
  \url{https://arxiv.org/abs/1903.05267}
\BIBentrySTDinterwordspacing

\bibitem{Ellis:2020lnc}
\BIBentryALTinterwordspacing
J.  Ellis, M.~A.~G.  Garcia, N.  Nagata, N.~D.  V., K.~A.  Olive, and S.
  Verner, ``{Building Models of Inflation in No-Scale Supergravity},'' 9 2020.
  \url{https://arxiv.org/abs/2009.01709}
\BIBentrySTDinterwordspacing

\bibitem{Ferrara:2019tmu}
\BIBentryALTinterwordspacing
S.  Ferrara, M.  Tournoy, and A.  Van~Proeyen, ``{de Sitter Conjectures in
  $N$=1 Supergravity},'' \emph{Fortsch. Phys.}, vol.~68, no.~2, p. 1900107,
  2020.  \url{https://arxiv.org/abs/1912.06626}
\BIBentrySTDinterwordspacing

\bibitem{Agrawal:2018own}
\BIBentryALTinterwordspacing
P.  Agrawal, G.  Obied, P.~J.  Steinhardt, and C.  Vafa, ``{On the Cosmological
  Implications of the String Swampland},'' \emph{Phys. Lett. B}, vol. 784, pp.
  271--276, 2018.  \url{https://arxiv.org/abs/1806.09718}
\BIBentrySTDinterwordspacing

\bibitem{Kallosh:2013yoa}
\BIBentryALTinterwordspacing
R.  Kallosh, A.  Linde, and D.  Roest, ``{Superconformal Inflationary
  $\alpha$-Attractors},'' \emph{JHEP}, vol.~11, p. 198, 2013.
  \url{https://arxiv.org/abs/1311.0472}
\BIBentrySTDinterwordspacing

\bibitem{Kallosh:2014rga}
\BIBentryALTinterwordspacing
------, ``{Large field inflation and double $\alpha$-attractors},''
  \emph{JHEP}, vol.~08, p. 052, 2014.  \url{https://arxiv.org/abs/1405.3646}
\BIBentrySTDinterwordspacing

\bibitem{Aghanim:2018eyx}
\BIBentryALTinterwordspacing
N.  Aghanim \emph{et~al.}, ``{Planck 2018 results. VI. Cosmological
  parameters},'' \emph{Astron. Astrophys.}, vol. 641, p.~A6, 2020.
  \url{https://arxiv.org/abs/1807.06209}
\BIBentrySTDinterwordspacing

\bibitem{Chiba:2012cb}
\BIBentryALTinterwordspacing
T.  Chiba, A.  De~Felice, and S.  Tsujikawa, ``{Observational constraints on
  quintessence: thawing, tracker, and scaling models},'' \emph{Phys. Rev. D},
  vol.~87, no.~8, p. 083505, 2013.  \url{https://arxiv.org/abs/1210.3859}
\BIBentrySTDinterwordspacing

\bibitem{Akrami_2018}
\BIBentryALTinterwordspacing
Y.  Akrami, R.  Kallosh, A.  Linde, and V.  Vardanyan, ``The landscape, the
  swampland and the era of precision cosmology,'' \emph{Fortschritte der
  Physik}, vol.~67, no. 1-2, p. 1800075, Nov 2018.
  \url{http://dx.doi.org/10.1002/prop.201800075}
\BIBentrySTDinterwordspacing

\bibitem{Copeland:1997et}
\BIBentryALTinterwordspacing
E.~J.  Copeland, A.~R.  Liddle, and D.  Wands, ``{Exponential potentials and
  cosmological scaling solutions},'' \emph{Phys. Rev. D}, vol.~57, pp.
  4686--4690, 1998.  \url{https://arxiv.org/abs/gr-qc/9711068}
\BIBentrySTDinterwordspacing

\bibitem{Brax:2006np}
\BIBentryALTinterwordspacing
P.  Brax and J.  Martin, ``{Moduli Fields as Quintessence and the Chameleon},''
  \emph{Phys. Lett. B}, vol. 647, pp. 320--329, 2007.
  \url{https://arxiv.org/abs/hep-th/0612208}
\BIBentrySTDinterwordspacing

\bibitem{Riess:2018byc}
\BIBentryALTinterwordspacing
A.~G.  Riess \emph{et~al.}, ``{Milky Way Cepheid Standards for Measuring Cosmic
  Distances and Application to Gaia DR2: Implications for the Hubble
  Constant},'' \emph{Astrophys. J.}, vol. 861, no.~2, p. 126, 2018.
  \url{https://arxiv.org/abs/1804.10655}
\BIBentrySTDinterwordspacing

\bibitem{Riess:2020fzl}
\BIBentryALTinterwordspacing
A.~G.  Riess, S.  Casertano, W.  Yuan, J.~B.  Bowers, L.  Macri, J.~C.  Zinn,
  and D.  Scolnic, ``{Cosmic Distances Calibrated to 1\% Precision with Gaia
  EDR3 Parallaxes and Hubble Space Telescope Photometry of 75 Milky Way
  Cepheids Confirm Tension with $\Lambda$CDM},'' \emph{Astrophys. J. Lett.},
  vol. 908, no.~1, p.~L6, 2021.  \url{https://arxiv.org/abs/2012.08534}
\BIBentrySTDinterwordspacing

\bibitem{Krishnan:2021dyb}
\BIBentryALTinterwordspacing
C.  Krishnan, R.  Mohayaee, E.~O.  Colg{\'a}in, M.~M.  Sheikh-Jabbari, and L.
  Yin, ``{Does Hubble Tension Signal a Breakdown in FLRW Cosmology?}'' 5 2021.
  \url{https://arxiv.org/abs/2105.09790}
\BIBentrySTDinterwordspacing

\bibitem{Banerjee_2021}
\BIBentryALTinterwordspacing
A.  Banerjee, H.  Cai, L.  Heisenberg, E.~O.  Colg{\'a}in, M.  Sheikh-Jabbari,
  and T.  Yang, ``Hubble sinks in the low-redshift swampland,'' \emph{Physical
  Review D}, vol. 103, no.~8, Apr 2021.
  \url{http://dx.doi.org/10.1103/PhysRevD.103.L081305}
\BIBentrySTDinterwordspacing

\bibitem{mortsell2021hubble}
\BIBentryALTinterwordspacing
E.  Mortsell, A.  Goobar, J.  Johansson, and S.  Dhawan, ``The hubble tension
  bites the dust: Sensitivity of the hubble constant determination to cepheid
  color calibration,'' 2021.  \url{https://arxiv.org/abs/2105.11461}
\BIBentrySTDinterwordspacing

\end{thebibliography}


\end{document}